\documentclass[%
 reprint,.
 amsmath,amssymb,
 aps,
]{revtex4-2}
\usepackage{physics}
\usepackage{graphicx}
\usepackage{dcolumn}
\usepackage{bm}
\usepackage{longtable}
\usepackage[toc,page]{appendix}

\begin{document}

\preprint{APS/123-QED}

\title{Understanding Excitations in $^{59,61}$Co, $^{59}$Ni}

\author{Samuel Ajayi}
 \email{soa19@fsu.edu}
\author{Vandana Tripathi}%
\email{vtripath@fsu.edu}
\author{E.~Rubino}
\author{Soumik Bhattacharya}
\author{L. T. Baby}
\author{R.~S.~Lubna}
\author{C.~Benetti}
\author{Catur Wibisono}
\author{MacMillan B. Wheeler}
\author{S.~L.~Tabor}
\affiliation{Department of Physics, Florida State University, Tallahassee, Florida 32306, USA}%


\author{Yutaka Utsuno}
\affiliation{
Advanced Science Research Center, Japan Atomic Energy Agency, Tokai, Ibaraki 319-1195, Japan
}%
\affiliation{
Center for Nuclear Study, University of Tokyo, Hongo, Bunkyo-ku, Tokyo 113-0033, Japan
}%
\author{Noritaka Shimizu}
\affiliation{Center for Computational Sciences, University of Tsukuba, 1-1-1
Tennodai, Tsukuba 305-8577, Japan}%
\author{J.~M.~Allmond}
\affiliation{
Oak Ridge National Laboratory, Physics Division, TN 37831-6371, USA
}%

\date{\today}

\begin{abstract}
High spin states in $^{59}$Co ($Z=27$), $^{59}$Ni ($Z=28$) and $^{61}$Co have been populated by the fusion 
evaporation reactions, $^{48}$Ti($^{14}$C, p2n)$^{59}$Co, $^{48}$Ti($^{14}$C, 3n)$^{59}$Ni, and 
$^{50}$Ti($^{14}$C, p2n)$^{61}$Co. The 9 MV tandem accelerator at the John D Fox Laboratory, Florida State 
University (FSU)  was used to accelerate the $^{14}$C beam and the de-exciting $\gamma$ rays were 
detected by the FSU detector array consisting 
of six High Purity Germanium (HPGe) clover detectors, and three single crystals. Directional correlation 
of the $\gamma$ rays de-exciting 
oriented states (DCO ratios) and polarization asymmetry measurements helped to establish spin and parities of 
the excited states whenever possible. The level scheme of $^{59}$Co has been expanded with the inclusion of 
positive parity states upto 31/2$^+$ at around 11 MeV. The $^{59}$Ni positive parity states 
known from previous study were verified with modifications to some of the spins and parities. On the other hand, 
the negative parity states were extended to 31/2 at an excitation energy of 12 MeV. 
No new transition was observed for $^{61}$Co, but one of the major bands has been 
reassigned as consisting of positive parity states by reason of this study. 
Excitations observed within the f$_{7/2}$, p$_{3/2}$, f$_{5/2}$ and p$_{1/2}$ orbitals, and also 
across the $N=40$ sub-shell closure into the g$_{9/2}$ orbital was 
established by comparison with large-scale shell model calculations for the three nuclei studied.
\end{abstract}
\maketitle



\section{\label{sec:level1} Introduction }
There has been a lot of interest in the study of nuclei around mass number A $\approx$ 60 in recent years. 
For nuclei in this mass region, the protons and neutrons both lie in the $fp$ shell, near the doubly magic, 
spherical nucleus $^{56}$Ni which can act as a natural core for understanding excitations. 
$^{56}$Ni has proton number and neutron number equal to 28 and therefore the nucleons fill up the f$_{7/2}$
orbital. Any addition or subtraction of nucleons to this spherical nucleus can have an effect on its structure, especially at higher 
excitation. Nuclei with valence nucleons in the p$_{3/2,1/2}$ and  f$_{5/2}$ orbitals, above the f$_{7/2}$ orbital, 
have the possibility of getting excited into the g$_{9/2}$ orbital which has been known to bring about collectivity.
The collective excitations that have been observed in this mass region consist of magnetic rotation in nearly 
spherical nuclei and also rotation due to a deformed nucleus. 
Therefore, nuclei in this region are perfect for the study of structural changes from spherical to deformed configurations as a function of energy and angular momentum.

Several studies have been performed to investigate the excited states in nuclei in the mass A $\approx$
60 region. Above the Z = 28 shell closure, the excited superdeformed states of $^{59}$Cu and $^{61}$Cu 
nuclei (Z = 29) have elucidated the evolution of nuclear shapes from spherical to deformed based on the 
$\nu$g$_{9/2}$ orbital \cite{59Cu, 61Cu}. Similarly, for nuclei with Z $\leq$ 28, the neutron g$_{9/2}$
induces deformation and the onset of collectivity as has been reported in a large number of nuclei in this 
mass region like $^{57\ensuremath{-}60}\mathrm{Mn}$ (Z = 25), $^{57}$Fe, $^{59, 60}$Fe (Z = 26), 
$^{61}$Co (Z = 27), and $^{58}$Ni (Z = 28) \cite{57-60Mn, 57Fe, 59_60_Fe, Anyageakaa, 58Ni}. 
Magnetic rotation bands which were first observed in the near-spherical Pb isotopes with A $\approx$ 200 \cite{198Pb, 199Pb}, 
have also been observed in lighter nuclei in the A$\approx$ 60 mass region, like $^{58}$Fe, $^{61}$Co, $^{62}$Co, $^{60}$Ni, 
$^{61}$Ni, and $^{62}$Cu \cite{58Fe, Anyageakaa, 62Co, 60Ni_Torres, Soumik, Lin61Ni, 62Cu}. 

To explain the experimental observations in this mass region both microscopic and macroscopic 
model have been employed. 
Shell model calculations have been quite successful in reproducing the low spin states in these
nuclei whereas  the particle rotor model and the cranked Nilsson-Strutinsky calculations have been used to interpret the rotational bands in many cases.
Zhao {\it et al.} used the self-consistent tilted axis cranking 
relativistic mean-field theory based on a point-coupling interaction to investigate magnetic rotation 
bands in $^{60}$Ni \cite{Zhao} while the Skyrme Hartree-Fock calculations was been used to describe the
rotational bands observed in $^{57}$Co \cite{57Co_Reviol, Dobacz}. 
Ref. \cite{meng2013progress} described the tilted axis cranking covariant density functional theory and 
its application for the magnetic rotation and antimagnetic rotation phenomen in different mass regions, 
while the Ref \cite{Lin61Ni} used the technique to investigate the rotational bands in $^{61}$Ni. 
Afanasjev {\it et. al.} used 
the cranked relativistic mean field theory and the configuration-dependent cranked Nilsson-Strutinsky approach to study superdeformed and
highly deformed rotational bands in the A = 60 mass region \cite{Afanasjev}. 


Magnetic dipole rotation observed in near spherical nuclei is usually characterized by bands
of strong $\mathrm{M1}$ transitions as opposed to the $\mathrm{E2}$ transitions which indicate rotation 
due to deformation. This phenomenon can be explained using the shears mechanism, where there is a coupling and gradual 
alignment of the spin of the protons and neutrons making up the total angular momentum of the levels, with the 
proton and neutron spin vectors as two blades of a shear
\cite{MagRot_82_84Rb, MagRot_112In, MagRot_Hubel}. 
The cross-over $\mathrm{E2}$ transitions in these bands are generally 
weak or sometimes not observed as documented in the previous studies on magnetic rotation \cite{MagRot_Hubel, Clark}. 
Magnetic transition probability, B(M1) is expected to decrease with an increase in the total 
angular momentum vector as the magnetic moment reduces with the closing of the shear blades. 

The isotopes, $^{59,61}$Co and $^{59}$Ni, which are the focus of this study all have protons 
occupying the $f_{7/2}$ orbitals (completely 
filled for $^{59}$Ni). The neutrons on the other hand lie in the p$_{3/2,1/2}$ and  f$_{5/2}$ orbitals, 
above the f$_{7/2}$ orbital, and have the possibility of getting excited into the g$_{9/2}$ orbital. The 
implication of this is that positive parity states can be generated at high spins, and since no experimental 
study on $^{59,61}$Co has reported this before, it will be interesting to probe these high-spin states to
observe the possibility. Also, given that magnetic rotation has been reported in $^{61}$Ni, it becomes 
necessary to investigate if it also exists in $^{59}$Ni which is just 2 neutrons less. The high-spin 
excitations give the perfect opportunity to investigate the structural changes from spherical to nearly 
deformed or deformed nucleus because their valence neutrons lie 
between the spherical closed shell nuclei and the deformation driving g$_{9/2}$ orbital.
Prior investigations of $^{59}$Co using fusion evaporation reactions have not particularly focused on studying such structural changes 
\cite{Warbuton,59Co_61Co_1970, 59Co_Haupt}. Previous studies of $^{59}$Ni and $^{61}$Co have however 
focused respectively on rotational bands and the role of the g$_{9/2}$ orbital in the development of 
collectivity \cite{Yu, Anyageakaa}. The experimental 
results from the high spin excitations will serve as a good testing ground of  large-scale shell model calculation in a valence space including the g$_{9/2}$ and d$_{5/2}$ orbitals.


\section{Experimental Details}

Two fusion-evaporation reaction experiments were performed at the John D Fox Laboratory, Florida State 
University (FSU) to populate and study the nuclei of interest. 
The beam used for both experiments is a long lived radioactive $^{14}$C beam which was accelerated 
to an energy of 43 MeV using the 9 MV tandem accelerator. 
The targets were thin unbacked foils of $^{48}$Ti and $^{50}$Ti with thickness around 500$\mu g/cm^2$, and 
highly enriched up to over 90\% for the particular isotope. The reaction $^{48}$Ti($^{14}$C, p2n)$^{59}$Co 
populated the high spin states of $^{59}$Co by the evaporation of a proton and two neutrons (p2n channel). 
High spin states of $^{59}$Ni were also populated using the same target in the reaction $^{48}$Ti($^{14}$C, 
3n)$^{59}$Ni by the evaporation of three neutrons from the compound nucleus (3n channel). The reaction 
$^{50}$Ti($^{14}$C, p2n)$^{61}$Co on the other hand, populated the high spin states of $^{61}$Co by the
evaporation of a proton and two neutrons from the compound nucleus formed (p2n channel).

The FSU $\gamma$-detector array consisting of six High Purity Germanium (HPGe) clover detectors, and three 
single crystal Germanium detectors was used for detecting the $\gamma$ rays from the excited states of the 
three nuclei. Three HPGe clover detectors were coupled to Bismuth Germanium Oxide (BGO) shields, an 
inorganic scintillator detector, which allows for Compton suppression. The detectors in the array were placed 
at 90$^{\circ}$, 45$^{\circ}$, and 135$^{\circ}$ with respect 
to the beam axis. This made the calculation for the Directional Correlation of Oriented States (DCO Ratio) 
possible. The energy and efficiency calibrations of the germanium detector array were performed using known 
calibrated $^{152}$Eu, $^{133}$Ba, and short-lived $^{56}$Co sources. The $^{56}$Co source was made at FSU 
using a proton beam. The spectra from the non 90$^{\circ}$ detectors were corrected for Doppler shift using 
the $\beta$ (v/c) value of recoiling nuclei and the detection angle.
The PIXIE digital data acquisition system was used to record the signals from the detectors and digitize for 
further analysis. For this experiment, the data was collected with a multiplicity setting greater than or equal to a 2-fold $\gamma$ coincidence.

\section{Analysis}

\begin{figure}
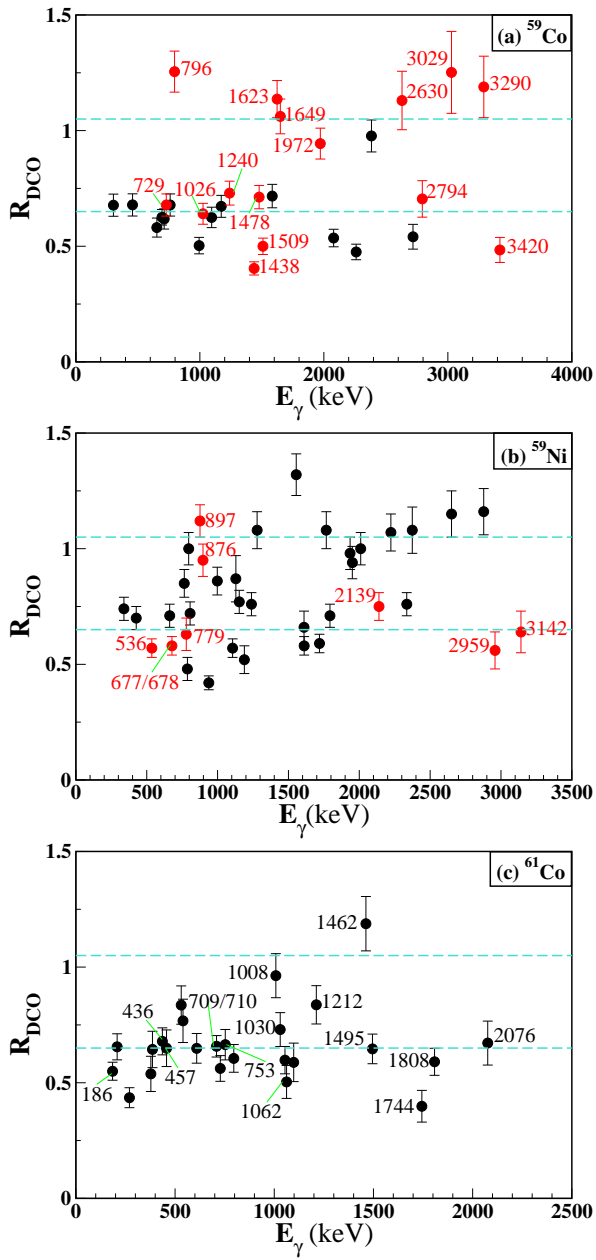

    \centering
    \includegraphics[clip,width=0.9\columnwidth]{59Co_DCORatio.eps}
    \includegraphics[clip,width=0.9\columnwidth]{DCO_59Ni.eps}
    \includegraphics[clip,width=0.9\columnwidth]{61Co_DCORatio.eps}
    \caption{(a) DCO ratio for transitions in $^{59}$Co
    (b) DCO ratio for transitions in $^{59}$Ni.
For both (a) and (b) points labeled in red are newly placed while those labeled in
    black are known transitions (c) DCO ratio for transitions 
    in $^{61}$Co. All existing transitions are labeled in black. Point 709/710 represents the two transitions 709 keV 
    and 710 keV which are both dipole in nature. For (a), (b), and (c), the transitions were from gates made 
    on stretched quadrupole transitions.} 
    \label{fig:DCO}
\end{figure}

The event mode data was built into $\gamma$ - $\gamma$ matrices using the Gnuscope software developed at FSU. 
This was used for $\gamma$ - $\gamma$ coincidence analysis to identify new 
$\gamma$ rays in coincidence with previously known transitions, which was then used to build up the level 
scheme as will be discussed.
Angle-dependent asymmetric matrices were also made for Directional Correlation of Oriented States 
(DCO Ratio) analysis. The DCO ratio ($R_{DCO}$) technique was used to determine the spin of transitions, which infers the spin change between the energy levels connected by the transition. This spin change by extension was used to assign the 
spin of the higher energy level between levels joined by the transition. 
This assumes that the spin of the lower energy level is already known.

Given that $\gamma_1$, $\gamma_2$ are two $\gamma$ peaks in coincidence, and $\theta_1$, $\theta_2$ are 
the angles of their detection,  90$^{\circ}$ and 135$^{\circ}$ in our case, the $R_{DCO}$ is given by 
\cite{KRANE_DCO}
\begin{equation}
    R_{DCO} = \frac{I^{\gamma_2}_{\theta_1}(Gate^{\gamma_1}_{\theta_2})}{I^{\gamma_2}_{\theta_2}(Gate^{\gamma_1}_{\theta_1})}
\end{equation}

where $I^{\gamma_2}_{\theta_1}(Gate^{\gamma_1}_{\theta_2})$ is the intensity of $\gamma_2$ determined from 
a spectrum in detectors at $\theta_1$ gated on $\gamma_1$ detected by detectors at $\theta_2$. 

For gates that were made on pure dipole transitions, if the $R_{DCO}$ is around $1\pm0.3$, then the
transition is dipole and if $R_{DCO}$ is around $1.85\pm0.35$, then it is a quadrupole transition. 
For gates that were made on pure quadrupole transitions,
if $R_{DCO}$ is around $0.55\pm0.15$, then the transition is dipole and if $R_{DCO}$ is around $1.05\pm0.25$, 
then the transition is quadrupole. Figure \ref{fig:DCO} shows the plot of $R_{DCO}$ with energy of 
$\gamma$ transitions in $^{59}$Co, $^{61}$Co, and $^{59}$Ni. The $\gamma$ transitions with known 
multipolarities were in good agreement with our analysis giving us the confidence to make predictions 
for the new transitions.

Measuring the polarization asymmetry of the emitted $\gamma$ rays helps in determining whether the $\gamma$ ray 
transitions are magnetic or electric in nature. With this information, we can assign parity to the new states 
identified by the $\gamma$ - $\gamma$ coincidence analysis, having assigned spin values 
using the DCO ratio technique. 
The four crystals in the clover detectors located at 90$^\circ$ to the beam axis served as Compton Scattering 
polarimeter which makes polarization asymmetry measurements possible. The value of the polarization asymmetry 
A is positive for electric transitions and negative for magnetic transitions, and is given by \cite{polarization}
\begin{equation}
   A = \frac{aN_\perp - N_\parallel}{aN_\perp + N_\parallel} 
   \label{eq:pol}
\end{equation}
$N_\perp$ and $N_\parallel$ are the numbers of $\gamma$ rays Compton scattered in the perpendicular and 
parallel directions with respect to the beam axis. The factor, {\it a} is a correction term that is defined at A = 0 as

\begin{equation}
    a = \frac{N_\parallel}{N_\perp}
\end{equation}

\begin{figure}
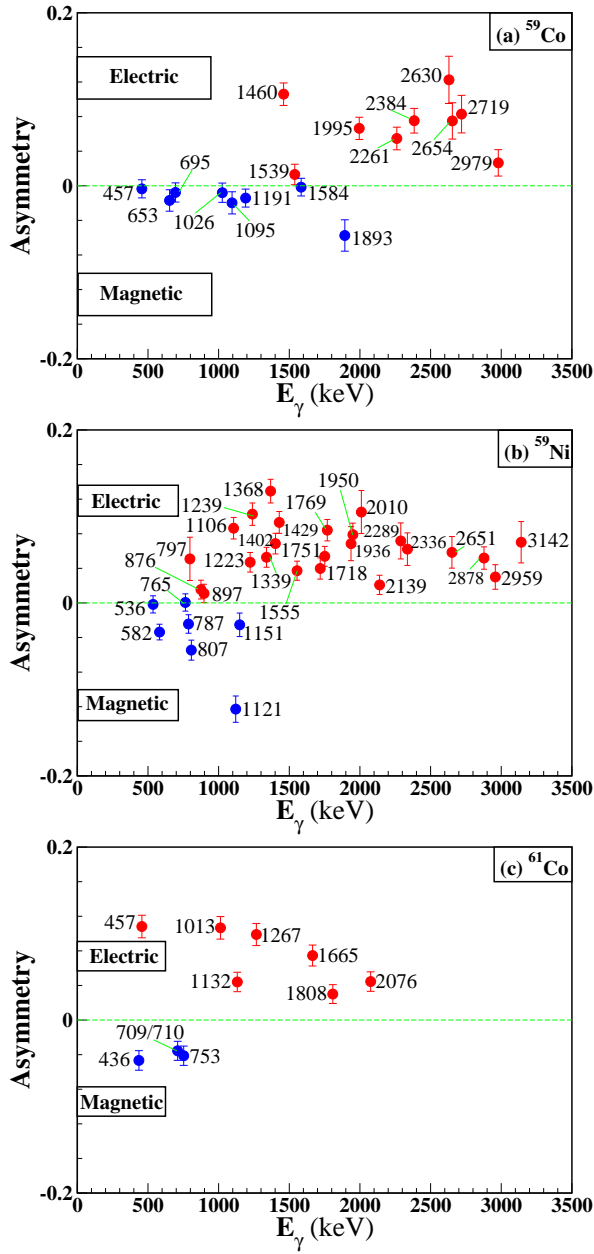

    \centering
    \includegraphics[clip,width=0.9\columnwidth]{59Co_Polarization.eps}
    \includegraphics[clip,width=0.9\columnwidth]{59Ni_Polarization.eps}
    \includegraphics[clip,width=0.9\columnwidth]{61Co_Polarization.eps}
    \caption{The plot of polarization asymmetry vs. energy of $\gamma$-ray for (a) $^{59}$Co, (b) $^{59}$Ni and (c) $^{61}$Co. Point 709/710 in (c) represents the two transitions 709 keV and 710 keV which are magnetic in nature. The points in red are electric transitions while the points in blue represent magnetic transitions.}
    \label{fig:asymmetry}
\end{figure}

The data was sorted into two hits for the clover detectors placed at 90$^\circ$ for 
polarization asymmetry; one parallel to the beam direction, and the other perpendicular to the beam direction. 
Using a $^{152}$Eu unpolarized source, we measured the factor {\it a}, for different energy which was then fitted
 with a linear equation to obtain the factor {\it a}, as a function of energy. The uncertainty in the measured 
{\it a} is given by $\Delta{\it a} = 0.0600$, and it applies to the intercept of the equation ($1.0372\pm0.0600$).

\begin{equation}
    a = -3.05 \times 10^{-5} E_\gamma + 1.0372, \Delta{\it a} = 0.0600
    \label{eq:littlea}
\end{equation}

For all the $\gamma$ transitions with good statistics, we calculated their 
polarization asymmetry as given in equations \eqref{eq:pol} and \eqref{eq:littlea}
and classified them as either electric or magnetic depending on the sign of A.
The Figure \ref{fig:asymmetry} shows the polarization asymmetry of the $\gamma$ transitions in
the 3 nuclei, and their classification as either electric or magnetic is indicated.
        
\section{Results}

The combination of $\gamma$ - $\gamma$ coincidence analysis, $R_{DCO}$ measurement, 
and polarization asymmetry measurements allowed us to build new expanded level schemes for 
the three nuclei in this study, $^{59}$Co, $^{59}$Ni, and $^{61}$Co which will
 be discussed in the following sections.

\subsection{Level Scheme of $^{59}$Co}

\begin{figure}
    \centering
    \includegraphics[width=0.9\linewidth]{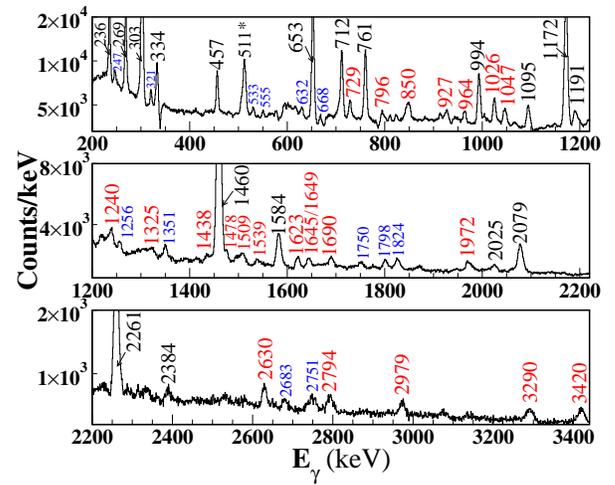}
    \caption{695-keV gate showing the $\gamma$ ray energy peaks in $^{59}$Co coincident with it. 
    All the peaks labeled in red were newly discovered in this study. The 511-keV peak due to pair 
    production is also shown; differentiated from other peaks with the * sign. The peaks labeled in blue are from 
    possible contaminants from other nuclei also produced in the reaction; mostly from $^{58}$Ni, $^{56}$Fe, and $^{53}$Cr.}
    \label{fig:Gate694}
\end{figure}

\begin{figure}
    \centering
    \includegraphics[width=0.9\linewidth] {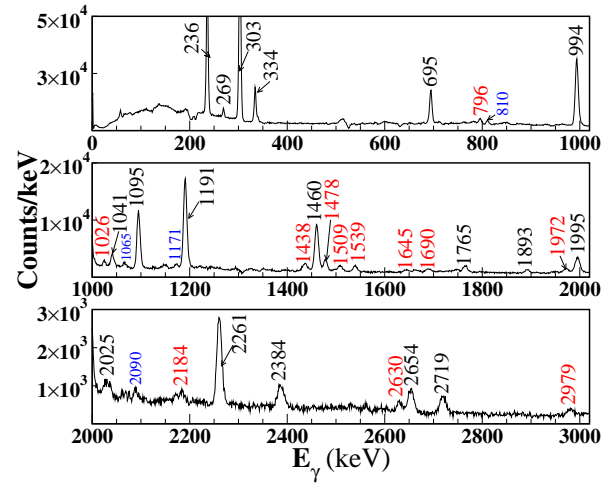}
    \caption{653-keV gate showing the $\gamma$ ray energy peaks in $^{59}$Co coincident with it. 
    All the peaks labeled in red were newly discovered in this study. The peaks labeled in blue are from 
    possible contaminants from other nuclei also produced in the reaction; mostly from $^{56}$Fe and $^{58}$Ni.}
    \label{fig:Gate653}
\end{figure}

\begin{figure*}
    \centering
    \includegraphics[width=1.0 \linewidth]{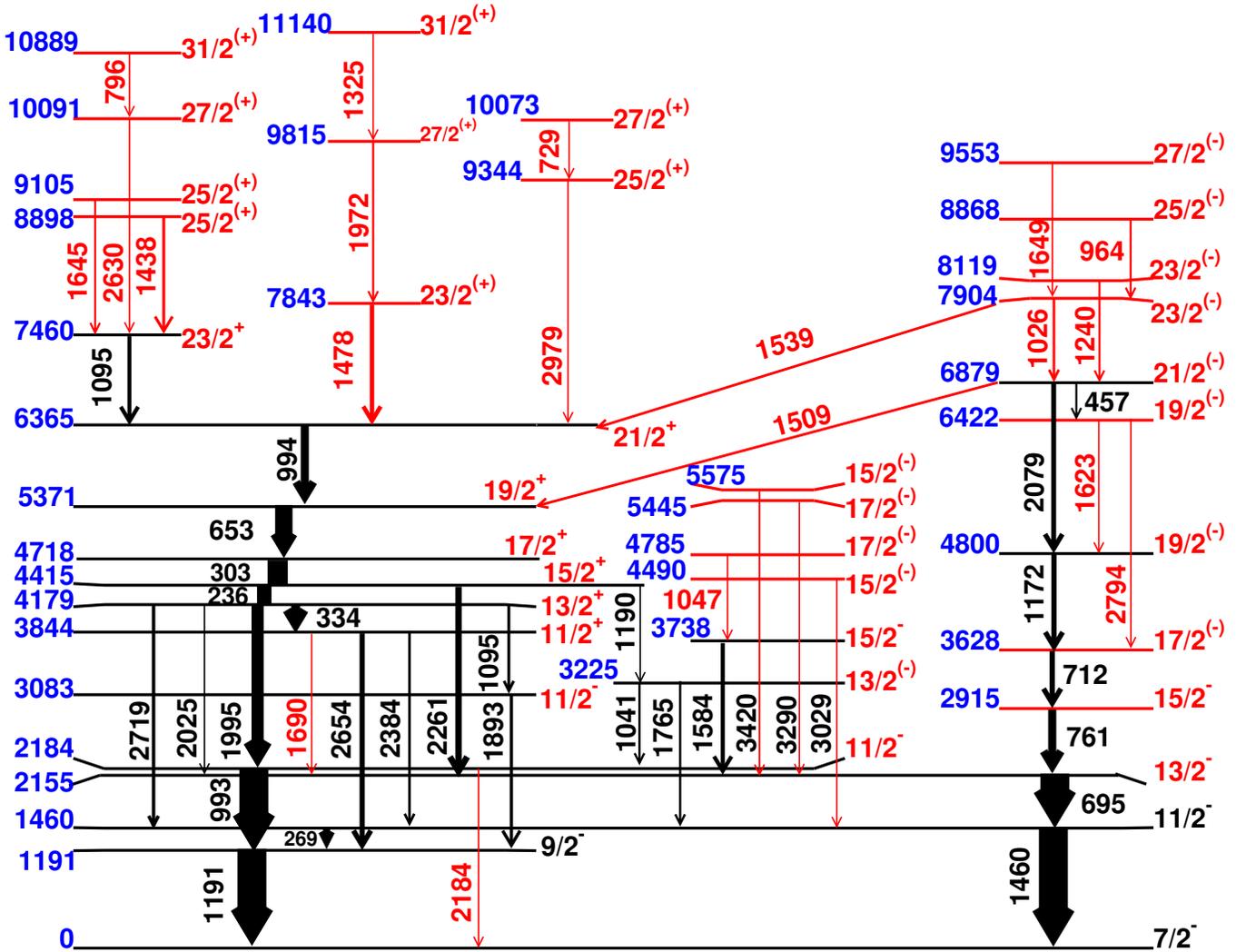}
    \caption{Level scheme of $^{59}$Co. All lines and labels in red are transitions and levels that were 
    newly discovered in this study, while those in black have been discovered in previous studies. 
    The thickness of the line arrows are approximately proportional to the intensity of the transitions. 
Spins and parities in parentheses are tentatively assigned based on 
expected patterns the level scheme since frequent parity change between consecutive levels is unlikely. The numbers in blue color are the energy level, distinguished from numbers in black 
which are the energies of the transitions.}
    \label{fig:59Co_level}
\end{figure*} 

In a previous study by Warburton {\it et al.} \cite{Warbuton} using the 
$^{48}$Ca($^{14}$N, 3n)$^{59}$Co reaction, a maximum spin of 23/2 at an excitation energy of about 
7.5 MeV was attained. In this current work, we have been able to confirm states predicted in the previous study and 
assigned parities to them. We have also extended the negative parity states to 27/2$^-$ at around 9.6 MeV, 
while identifying positive parity 
states upto a J$^{\pi}$ = 31/2$^+$ which had not been observed in any previous study of $^{59}$Co. 
The multipolarity of the transitions in this study based on the value of the 
$R_{DCO}$ and polarization asymmetry are given in Table \ref{tab:59Co}. Figures \ref{fig:DCO}(a) and 
\ref{fig:asymmetry}(a) display the numbers for $^{59}$Co. The $R_{DCO}$ value for the new transitions as seen 
in Figure \ref{fig:DCO}(a) is based on the fact that the transition gated upon is a quadrupole transition. 

Figures \ref{fig:Gate694} and \ref{fig:Gate653} show the coincident $\gamma$ ray peaks when 
gates were defined on the 695-keV and the 653-keV transitions. These two gates show most of the 
transitions already known (labeled in black), and the new ones observed in this study 
(labeled in red). The peaks labeled in blue in the figures are mostly from other contaminants produced in 
this reaction. There are also a few peaks that might be in $^{59}$Co but which we were unable to place in the level scheme. The expanded level scheme of $^{59}$Co can be seen in Figure \ref{fig:59Co_level}.

We have made an adjustment to two energy levels from the published level scheme in Ref \cite{Warbuton}. 
The 3326-keV and 4087-keV levels have now been replaced with the 2915-keV and 3628-keV levels. 
This is driven by the fact that the observed intensities for the $\gamma$ rays feeding the levels 
did not support the arrangement in the previous study. There are 
 closeby 1168-keV and 1177-keV $\gamma$ transitions observed in $^{56}$Mn and $^{60}$Co respectively, which 
 influence the intensity of the 1172-keV $\gamma$ transition observed in $^{59}$Co.
These nuclei are all produced in the reactions $^{14}$C + $^{48}$Ti (this current study) and 
$^{14}$N + $^{48}$Ca (previous study by study by Warburton {\it et al.} \cite{Warbuton, Warburton2}) according to 
PACE calculations \cite{PACE, PACE_Gavron}. The intensity of the 1172-keV 
transition in $^{59}$Co according to this study was used to determine its placement, and 
hence changed from where it was placed in the level scheme of 
Ref \cite{Warbuton} as seen in Figure \ref{fig:59Co_level}.

A major addition from the current work for $^{59}$Co, is the identification of 
positive parity states which form a cascade. The transitions at 2654-, 1995- and 
2261-keV  were already known from previous studies and their placement could be verified in the current 
study. From the present $R_{DCO}$ values these transitions were confirmed to be dipole transitions, 
but further, the polarization asymmetry measurements suggest that they are of 
electric nature, making them $\mathrm{E1}$ transitions. 
This implies that they connect states with opposite parities. With the parities of the lower levels being 
negative we can conclude that the 2654-keV transition links the 11/2$^+$ to the 9/2$^-$ state, 
the 1995-keV transition links 13/2$^+$ to 11/2$^-$ state and the 2261-keV transition links the 15/2$^+$ 
to the 13/2$^-$ state. 
Beyond the 13/2$^+$ state, a series of relatively strong $\mathrm{M1}$ transitions were observed connecting the 
positive parity states with no measurable crossover $\mathrm{E2}$ transitions. 
The multipolarities concluded for all the transitions based on $R_{DCO}$ values and polarization asymmetry are 
listed in Table \ref{tab:59Co} with some plotted in Figures \ref{fig:DCO}(a) and \ref{fig:asymmetry}(a). Parities in 
parenthesis as seen in Table \ref{tab:59Co} and Figure \ref{fig:59Co_level} are tentatively assigned based on 
expected patterns of transitions between consecutive levels, knowing that a frequent parity change between the levels is unlikely.

In all, 22 new transitions were observed which resulted in 20 new energy levels. 
Nine of these new energy levels are of positive parity, in addition seven existing energy levels have now  
been identified as positive parity states.

\subsection{Level Scheme of $^{59}$Ni}

\begin{figure}
    \centering
    \includegraphics[width=0.9\linewidth]{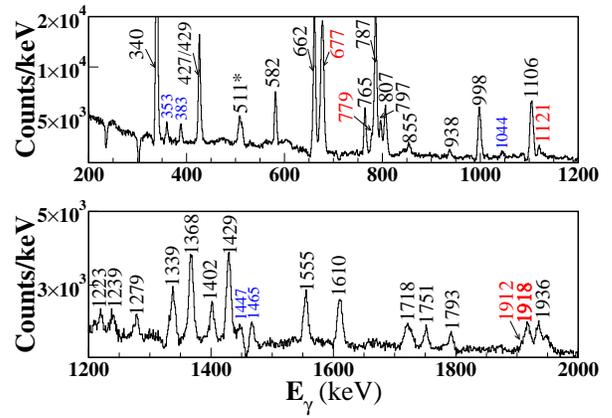}
    \caption{Gate on 678-keV transition in $^{59}$Ni. Another new 677-keV transition can be seen here. 
    All the peaks labeled in red are some of the new transitions just observed in this study, including 
    the 1918-keV transition. The 511-keV peak due to pair production is also shown; differentiated from 
    other peaks with the * sign. The peaks labeled in blue are from 
    possible contaminants from other nuclei also produced in the reaction; mostly from $^{58}$Co, $^{60}$Ni, and $^{56}$Fe.}
    \label{fig:678}
\end{figure}

\begin{figure}
    \centering
    \includegraphics[width=0.9\linewidth]{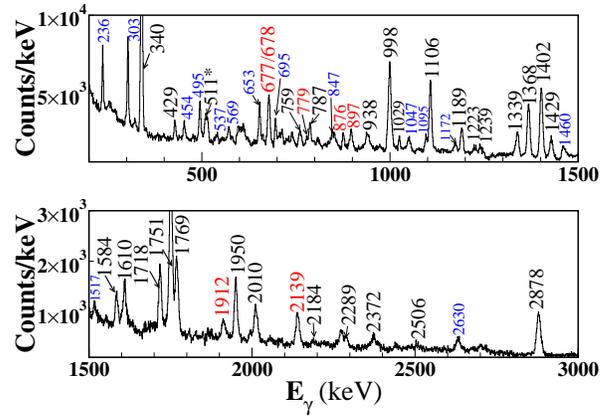}
    \caption{Gate on 797-keV transition in $^{59}$Ni. All the new transitions are labeled in red, 
    including the 1912-keV transition. The 511-keV peak due to pair production is also shown; 
    differentiated from other peaks with the * sign. The peaks labeled in blue are from 
    possible contaminants from other nuclei also produced in the reaction; mostly from $^{59}$Co, $^{58}$Ni, and $^{60}$Ni.}
    \label{fig:797}
\end{figure}

\begin{figure}
    \centering
    \includegraphics[width=0.9\linewidth]{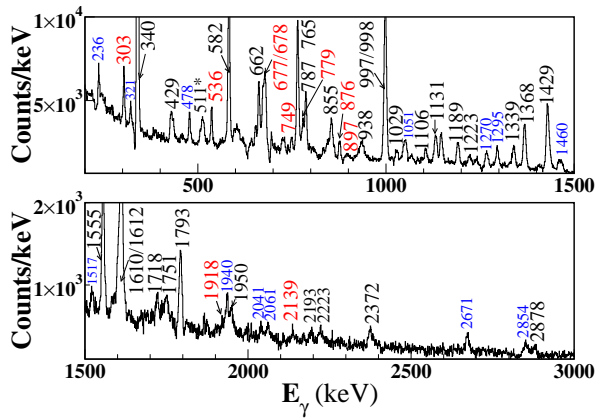}
    \caption{Gate on 807-keV transition in $^{59}$Ni. All the new transitions from the two negative bands 
    show up here except for the 1918-keV transition. The 511-keV peak due to pair production is also shown; 
    differentiated from other peaks with the * sign. The peaks labeled in blue are from 
    possible contaminants from other nuclei also produced in the reaction; mostly from $^{58}$Co, $^{60}$Ni, $^{58}$Ni, $^{59}$Co, and $^{55}$Fe.}
    \label{fig:807}
\end{figure}

\begin{figure*}
    \centering
    \includegraphics[width=1 \linewidth]{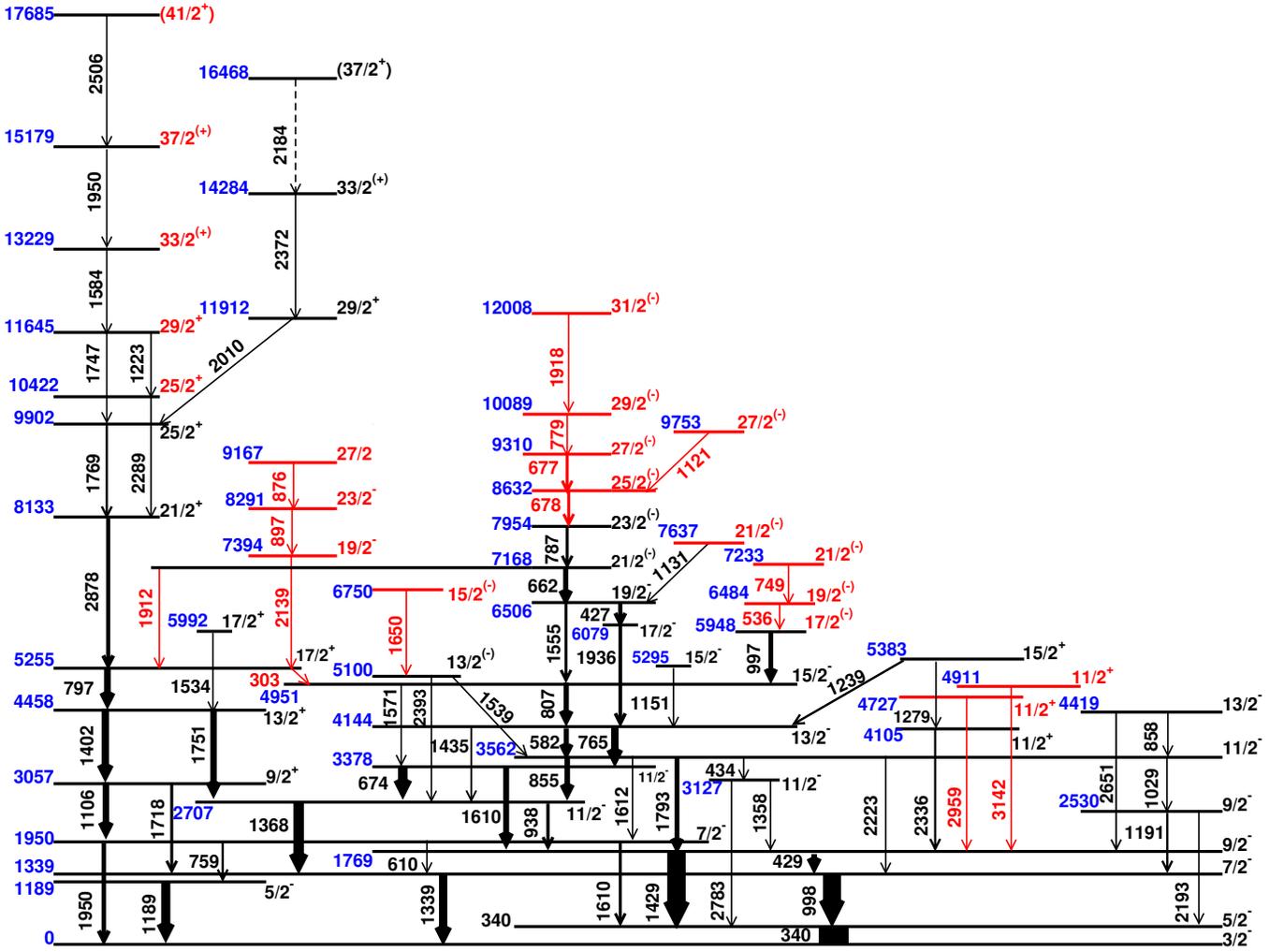}
    \caption{Level scheme of $^{59}$Ni showing the existing transitions and the new transitions observed 
    in this study. All the energy levels in red are the newly established levels in this study. 
    The thickness of the line arrows are approximately proportional to the intensity of the transitions. 
Spins and parities in parentheses are those tentatively assigned based on 
expected patterns the level scheme since frequent parity change between consecutive levels is unlikely. The numbers in blue color are the energy level, separated from numbers in black 
which are the energies of the transitions}
    \label{fig:59Ni_level}
\end{figure*}

In an earlier study by Juutinen {\it et al.} \cite{JUUTINEN} in 1989, using the reaction 
$^{58}$Ni($^{3}$He, 2p)$^{59}$Ni, negative parity states were observed upto an excitation energy of 
7.9 MeV with likely spin of 19/2$^-$ or 21/2$^-$. Positive parity states were also identified in the 
study, and the band built on the 9/2$^+$ state at 3057 keV was expanded to an energy level of 
around 6 MeV, though definitive spin assignments were not made. 

In a more recent study by Yu {\it et al.} \cite{Yu}, using the reaction 
$^{40}$Ca($^{29}$Si, 2$\alpha$2p)$^{59}$Ni, four rotational bands were established with the 
highest spin observed to be 43/2. The two strongest bands (1 and 2 in Ref.\cite{Yu}) 
are proposed to be generated by 
exciting one neutron and one neutron + one proton to the $g_{9/2}$ orbital respectively. 
They also indicated 
that the transition quadrupole moments of these bands decrease with spin suggestive of band termination. 

In the present study, new transitions were established which were instrumental in the extension 
of the existing band of negative-parity states to around 12 MeV with spin 31/2$^-$ which was earlier known 
only till 23/2 at an energy of 7954 keV from Ref.~\cite{JUUTINEN}. 
There is a new doublet of $\gamma$-ray 
transitions, 677-keV and 678-keV which appear in this band and can be seen in 
Figure \ref{fig:678} where the
gate made on the 678-keV transition clearly shows the other peak measured as 677-keV. 
The intensity of the 677-keV implies the two transitions when arranged in the level scheme 
should be in close proximity. Figure \ref{fig:678} also shows some of the other new transitions: 779-keV, 
1121-keV and 1918-keV associated with this band of negative parity states.
There is also a connection observed to the 17/2$^+$ state at 5255 keV via a new 1912-keV transition. 
 Though this transition is close in value to other new transition, 1918-keV which exists in the band, 
the gate made on the 797-keV transition as seen in Figure \ref{fig:797} validates the placement of the 
1912-keV transition. From the 807-keV gate shown in Figure \ref{fig:807}, we could see some of the existing 
transitions and some of the new transitions placed in the $^{59}$Ni level scheme. The expanded level scheme 
of $^{59}$Ni as a result of this work is shown in Figure~\ref{fig:59Ni_level} where red indicates new additions to the $\gamma$ transitions.

A new sequence of negative-parity states is also proposed and built on the 17/2$^+$ state at 5255 keV up to an
energy of 9167 keV with a spin of 27/2. 
The new transition 2139 keV was clearly established to be an $\mathrm{E1}$ transition according to Figures 
\ref{fig:DCO}(b) and \ref{fig:asymmetry}(b) and it connects the negative-parity state 19/2$^-$ to the 17/2$^+$ 
state. The 897-keV transition placed above the 
2139-keV transition to link the 23/2$^-$ to 19/2$^-$ was further established as an $\mathrm{E2}$ transition. We 
could not estimate the polarization asymmetry of the last transition 876-keV in the 
sequence though we could confirm it to be quadrupole in nature from the $R_{DCO}$ 
value. Two new transitions were also added to the band of negative-parity states that terminated at 5.9 MeV in
the previous study by Yu {\it et al.,} \cite{Yu}. No spin or parity was assigned to the 5.9-MeV state then,
but we have been able to assign a 17/2$^{(-)}$ to it in this study. We assumed that the dipole transition 997 
keV is most likely to be magnetic in nature, in line with the multipolarity of the 807-keV, 582-keV, and 536-keV 
transitions placed below and above it. The band was extended to 7.2-MeV at 21/2$^{(-)}$ by the 536-keV and 749-keV transitions. 
 
The sequence of positive parity states seen in the level scheme of $^{59}$Ni (Figure~\ref{fig:59Ni_level})
is analogous to band 1 and band 2 in Ref.~\cite{Yu} built on the 9/2$^+$ state, though 
we found differences in some of the proposed spins and parities. 
The three transitions, 1106-keV, 1718-keV and 1751-keV had been 
identified as dipole transitions in previous studies. Based on our polarization asymmetry 
measurements we further confirmed them to be electric in nature and assigned them as $\mathrm{E1}$ transitions. 
With that the spin-parities of the 9/2$^+$ and 13/2$^+$
are confirmed. Further up the 979-, 2878- and 1769-keV transitions build the band to 25/2$^+$ 
at 9902 keV. We firmly established 2878 keV and 1769 keV as $\mathrm{E2}$ transitions. 
Beyond that we see disagreements with the level structure proposed in Ref.~\cite{Yu}. In the gate of 
1106 keV we could see the 1747 keV transitions and its $R_{DCO}$ suggests it to be 
quadrupole and not dipole as in Ref.~\cite{Yu}. Similarly, the 2289-keV transition 
is established as a $\mathrm{E2}$ transition opposite to Ref.~\cite{Yu} where it was 
thought to be an $\mathrm{E1}$ transition though no polarization measurements are reported there. 
The $R_{DCO}$ value for 2289-keV transitions was ascertained from two gates (both confirmed $\mathrm{E2}$)
namely 797 keV and 2878 keV making it a $\mathrm{E2}$ transition.
We further confirmed that the 2010-keV and 1223-keV transitions are $\mathrm{E2}$ transitions. 
The 2372-keV transition was shown to be a quadrupole transition, though its asymmetry could not be ascertained. Using all this information, we established a sequence of $\mathrm{E2}$ 
transitions connecting all positive parity states extending to 17.7 MeV with spin 41/2$^+$. The multipolarities 
concluded for all the transitions based on $R_{DCO}$ values and polarization asymmetry are listed in Table \ref{tab:59Ni}. Parities (and spins) 
in parenthesis as seen in Table \ref{tab:59Ni} and Figure \ref{fig:59Ni_level} are tentatively assigned based on expected patterns of 
transitions between consecutive levels, and knowing that a frequent parity change between the levels is unlikely.

Overall, we observed 15 new transitions, and these in addition to some rearrangements brought about 
14 new energy levels in the level scheme of $^{59}$Ni.

\subsection{Level Scheme of $^{61}$Co}

\begin{figure}
    \centering
    \includegraphics[width=0.9\linewidth]{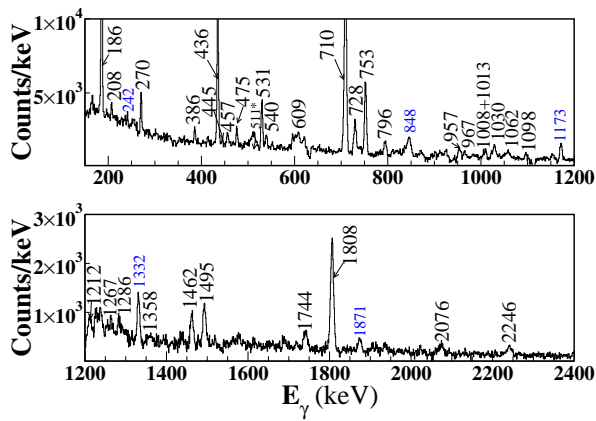}
    \caption{Gate on the 1665-keV transition showing some of the already established transitions in $^{61}$Co. 
    No new transition was observed in this study, so all the peaks are labeled in black. The 511-keV peak 
    due to pair production is also shown; differentiated from other peaks with the * sign. 
    The peaks labeled in blue are from 
    possible contaminants from other nuclei also produced in the reaction; mostly from $^{60}$Ni.}
    \label{fig:1665}
\end{figure}

\begin{figure*}
    \centering
    \includegraphics[width=1.0 \linewidth]{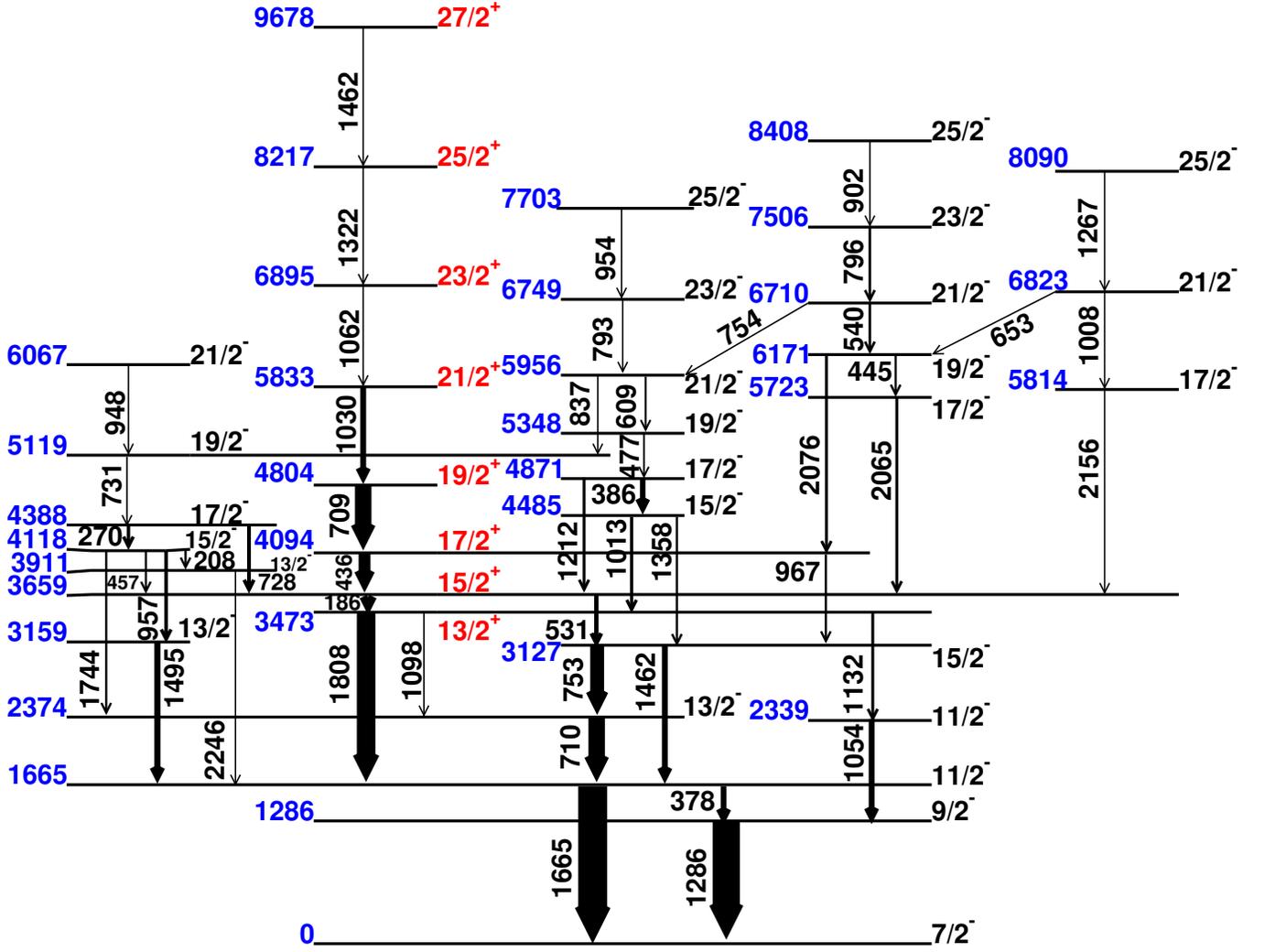}
    \caption{Level scheme of $^{61}$Co confirming the existing transitions and energy levels. 
    The spin and parity shown in red are the new important features added to this level scheme. 
    The thickness of the line arrows are approximately proportional to the intensity 
    of the $\gamma$-ray transitions. The numbers in blue color are the energy level, separated from numbers in black 
    which are the energies of the transitions}
    \label{fig:61Co_level}
\end{figure*}

In a previous detailed study by Ayangeakaa  {\it et al.} \cite{Anyageakaa}, the multinucleon transfer reaction, 
$^{26}$Mg($^{48}$Ca, 2$\alpha$4np$\gamma$)$^{61}$Co in inverse kinematics was used to study the 
excited states of $^{61}$Co. Six bands were identified as shown in the level scheme 
from that study, and all levels were assigned negative parity. 
In the current work, following the reaction $^{50}$Ti($^{14}$C, 2np)$^{61}$Co, we were able to observe 
five of the six bands previously identified. We were able to verify most of the transitions though we could not add any new $\gamma$-ray transitions. 
Figure~\ref{fig:1665} shows some of the transitions seen when a gate was made on the 1665-keV transition. 
Only two of the bands in this current study terminated on similar energy levels as in Ref.~\cite{Anyageakaa}. 
 This is because
of the higher $\gamma$-ray detection efficiency of Gammasphere used in Ref. \cite{Anyageakaa} compared to the 
FSU $\gamma$ array of primarily 6 clovers. Conversely the use of Clover detectors as 
Compton polarimeters in the current study allowed us to examine the parities of the levels. The level scheme as ascertained in this work is shown in 
Figure~\ref{fig:61Co_level} where the bands are referred to as 1 to 5 going from left to right.

Consistent with the previous study, only in the fifth band starting with the 17/2$^-$ state 
quadrupole transitions were observed connecting the exited states; all the other bands have 
a series of dipole transitions. We were able to clearly show that the 1267-keV has a $\mathrm{E2}$ 
multipolarity, though for the 1008 keV we could only get a $R_{DCO}$ value (Figure \ref{fig:DCO}(c)) 
suggesting it to be quadrupole in nature. This band in Ref.~\cite{Anyageakaa} had 5 more 
transitions going upto a spin of 41/2  which we could not observe.
But with the additional information from the current experiment, we can confirm that this band 
consists of $\mathrm{E2}$ transitions.

As mentioned before, the inclusion of polarization measurements in this study to determine which 
transition is electric or magnetic in nature brought about a major change to the level scheme of $^{61}$Co. 
The strong 1808-keV transition (see Fig.~\ref{fig:1665}) connecting the states at 3473 keV and 1665 keV
 has now been confirmed to be dipole in nature as seen in the Figure \ref{fig:DCO}(c). 
The polarization asymmetry measurement suggested that this transition is electric in nature as shown in 
Figure \ref{fig:asymmetry}. This ascribes 1808-keV as an $\mathrm{E1}$ transition which would link states of 
opposite parity. Hence it can be confirmed that the 1808-keV transition starts from a 13/2$^+$ state at 
3473 keV feeding the 11/2$^-$ state at 1665 keV. The transition at 1132 keV was also confirmed to be dipole 
and electric in nature, thus making it an $\mathrm{E1}$ transition. This served as an 
additional confirmation of the 13/2$^+$ state in this study as it ends on the 11/2$^-$ level. 
The 2076-keV transition which links bands 2 and 4 is also shown to be of $\mathrm{E1}$ in nature, so we 
could firmly determine the 4094 level as positive parity. We adopted the parity assignment of the 6171-keV level from Ref \cite{Anyageakaa} as negative. As for the 186-keV transition, we could only measure its $R_{DCO}$ and not its polarization asymmetry, because being low in energy, the $\gamma$ ray doesn't scatter, thereby making polarization measurement impossible. Even though we only know the 186-keV transition is dipole in nature, we could 
indirectly infer its multipolarity, knowing that the 436-keV transition which is an $\mathrm{M1}$
transition will link the 17/2$^+$ state to the 15/2$^+$ state at 3659-keV.
With this information, we can state that the transitions above the 3473 keV level, namely 
186-, 436- and 709 keV are $\mathrm{M1}$ in nature. Beyond that, it was difficult to get both $R_{DCO}$ and 
polarization asymmetry for all the transitions in the band because of low statistics and in the case of 1462 keV 
being a doublet. Details about the transitions, including their $R_{DCO}$ values and polarization asymmetry are 
listed in Table \ref{tab:61Co}.
In view of all the experimental observables from this work and Ref.~\cite{Anyageakaa}, we 
can confidently state that band 2 starting at the 13/2$^+$ state (3473 keV) is a sequence of $\mathrm{M1}$ 
transitions connecting states with positive parity. It also agrees very well with shell model calculations 
as will be discussed next.

\section{Discussion}

The odd-A nuclei studied in this work all have their valence nucleons occupying the $fp$ shell according 
to the simple shell model picture. The Co isotopes, $^{59}$Co and $^{61}$Co have their unpaired 
27$^{th}$ proton in the f$_{7/2}$ orbital, leading to a ground state spin/parity of 7/2$^-$. 
$^{59}$Ni on the other hand has a closed shell for protons but its unpaired valence neutron lies in the 
p$_{3/2}$  orbital, so its ground state is 3/2$^-$. Since the parity of the ground state energy level 
is negative, the confirmation of positive parity of some of the 
high spin states correspond to excitations across the $N=40$ sub-shell closure into the positive parity 
$g_{9/2}$ orbital. The excitations into the $g_{9/2}$ orbitals can generate high spin and are also 
responsible for generating collective motion.

The shell model calculations to interpret the data were performed utilizing the M-scheme code KSHELL on the Oakbridge-CX 
supercomputer at the University of Tokyo \cite{SHIMIZU}. Both negative and positive parity states 
were generated from the shell model calculation. The model space was taken as $fp$ shell for the 
negative parity states and the GXPF1Br interaction was used for that. For the positive parity states however, the model space 
was taken as the $fp$ shell, 0g$_{9/2}$, and 1d$_{5/2}$ orbits, restricting one neutron excitation to 
the 0g$_{9/2}$, and 1d$_{5/2}$ orbits. In addition, up to 6-particle 6-hole excitation from the 0f$_{7/2}$ 
orbit is allowed for the case of $^{59}$Ni and $^{59}$Co. The GXPF1Br+$V_{MU}$ interaction was used 
for the full calculation \cite{VMU_Interraction}. The predictions of the shell model calculation for the 
three nuclei in consideration align well with the experimental results. The root-mean-square (RMS) difference 
between the experimental results and the theoretical calculation averages below 200 keV. Table \ref{tab:Exp_Theo_Table} and Figures \ref{fig:59Co_ExpThe}, 
\ref{fig:59Ni_ExpThe}, \ref{fig:61Co_ExpThe} show the comparison between the calculation and experiment.
For the figures, the asterisks in red represent the experimental results and the squares in blue 
are the results from the shell model calculation for the yrast states. Figures \ref{fig:59Co_ExpThe} (a), 
\ref{fig:59Ni_ExpThe} (a) and \ref{fig:61Co_ExpThe} (a) additionally include the yrare 
negative-parity states 
({\it i.e.} 2nd excited state of each spin), with the experimental values represented by green circles, 
and the calculation represented by the green line. 
For the states where there are existing experimental values for the mean lifetime, we estimated the theoretical mean 
lifetime using the predicted $B(M1)$ values and found them to be in quite good agreement.
The details of the calculations will be discussed 
in greater detail below, in relation to the experimental results for each nuclide.

 \subsection{$^{59}$Co}
 
 \begin{figure}
    \centering
     \includegraphics[clip,width=\columnwidth]{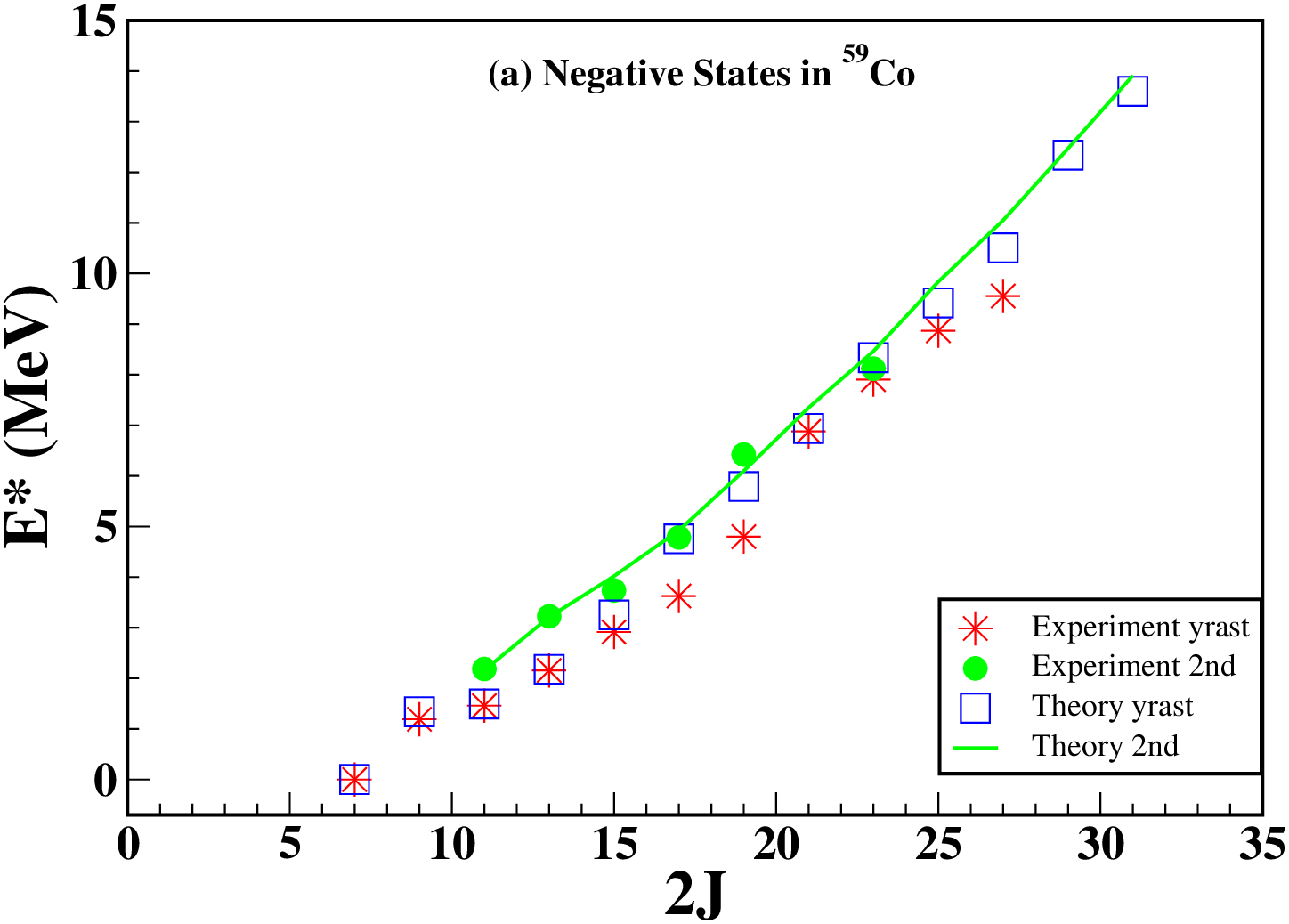}
    \includegraphics[clip,width=\columnwidth]{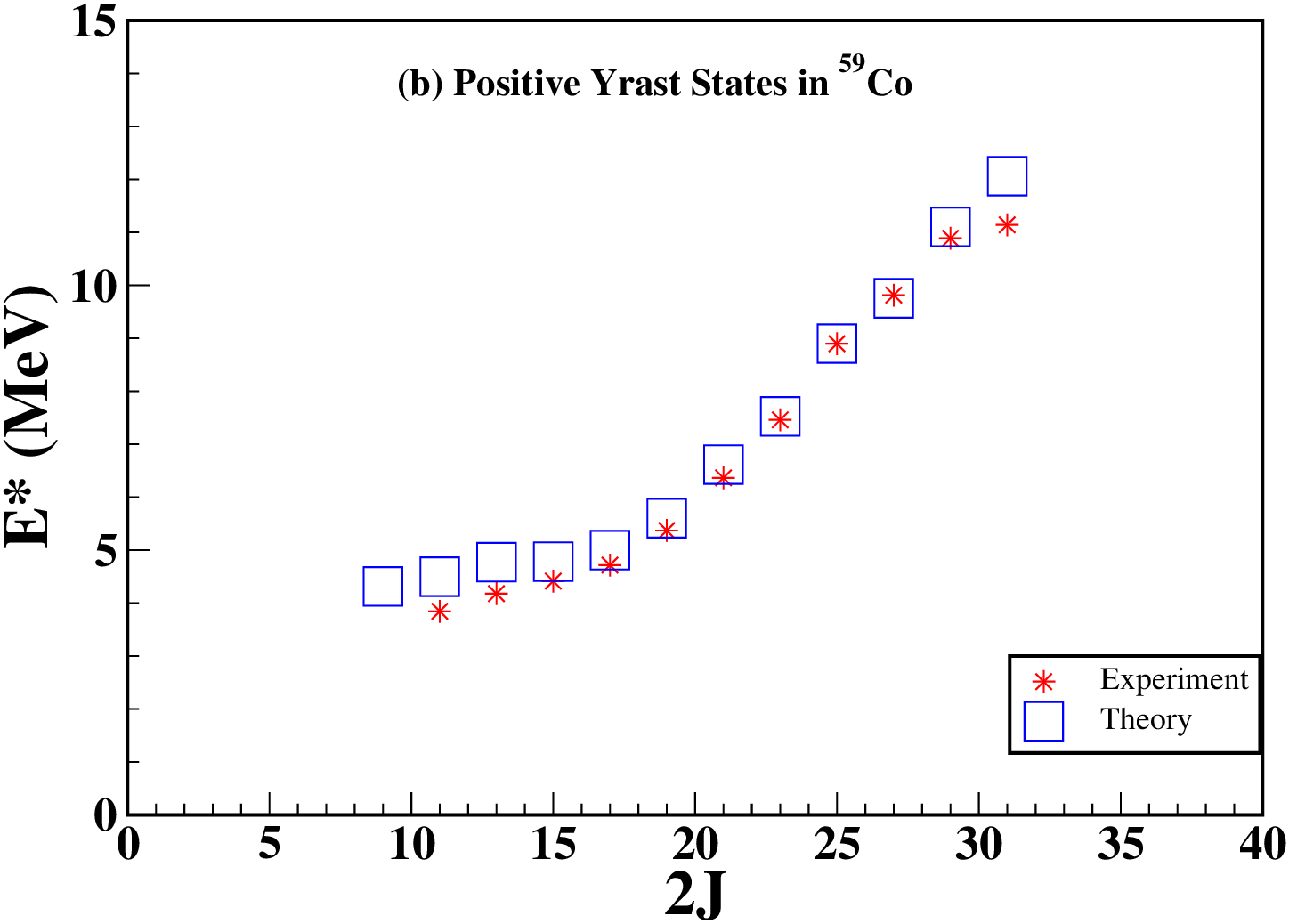}
    \caption{Comparison of theoretical shell model calculation with experimental results in $^{59}$Co. 
    The red star symbol represents experimental values while the blue box symbol represents values 
    from the theoretical shell model calculation. In (a), the ``experiment 2nd" and the ``theory 2nd" are the second states (near y-rast, also called yrare states), 
and are represented with green color; circles for experimental values, and line for the theoretical shell model calculation.}
    \label{fig:59Co_ExpThe}
\end{figure}

 $^{59}$Co has 27 protons and 32 neutrons, leading to all its valence nucleons in the $fp$ shell. 
 With 7 protons outside the closed $sd$ shell at $Z = 20$, we expect an unpaired proton in the 0f$_{7/2}$ orbit. 
 The 12 neutrons on the other hand outside the closed sd shell are all paired and not contribute to the
 ground state spin value expected to be 7/2$^-$ because of the unpaired proton. Of the 12 neutrons, 
 8 neutrons fill the 0f$_{7/2}$ orbit, while the remaining 4 neutrons are expected to fill the 
1p$_{3/2}$ orbit. However, as 1p$_{3/2}$ and 0f$_{5/2}$ levels are close in energy, both levels 
share the 4 neutrons in the ground state as per the calculations performed. This ground state 
configuration is verified by picturing $^{60}$Co as $^{59}$Co + n in the $^{59}$Co (d,p) reaction 
studied by Roy {et. al.,} to produce $^{60}$Co \cite{Roy}. If the four neutrons outside the closed 
0f$_{7/2}$ shell filled up the 1p$_{3/2}$ orbit, then the (d,p) reaction leading to the ground state of 
$^{60}$Co should be characterised by an $\ell$ = 3 transition (coming from 
the 0f$_{5/2}$ level). But all (d,p) reactions leading to the ground state and the isomeric state at 
60 keV have $\ell =1$ (coming from the 1p$_{3/2}$ level), therefore pointing to the fact that there is a 
vacancy in the 1p$_{3/2}$ orbit in $^{59}$Co to accommodate a neutron. We conclude therefore that in 
the ground state, $^{59}$Co likely has 2 valence neutrons each in the 1p$_{3/2}$ and 0f$_{5/2}$ levels 
consistent with the shell model calculations.  
 
 The excitations at low spin to negative parity states are likely single-particle excitations within the 
 $fp$ shell,  {\it i.e.}  $0p0h$ excitations. The shell model calculations for the $0p0h$ negative-parity states 
 align well with the experimental result up till the 27/2$^-$ level as seen in Figure~\ref{fig:59Co_ExpThe} and Table~\ref{tab:Exp_Theo_Table}. 
 The positive parity states arise from $1p1h$ excitations where a nucleon is excited to the g$_{9/2}$ 
 orbit leaving an even number of neutrons in the $fp$ shell. 
 From Figure \ref{fig:59Co_ExpThe}, we can see a good agreement between the shell model 
 calculation and the experimental result. The 9/2$^+$ state predicted by the calculation as the lowest 
 positive parity state was not observed in the experiment because it is not yrast.
 
 The positive parity states in $^{59}$Co form a regular pattern with a series of $\mathrm{M1}$ transitions 
 from the 13/2$^+$ state upwards to the 23/2$^+$ state. Experimental values for the mean lifetime of the states 
17/2$^+$, 19/2$^+$, and 21/2$^+$ have been given by Ref \cite{Warbuton} as 1.1(4) ps, 0.10(5) ps, and $\textless$ 0.20 ps respectively.
The B(M1) (down) from the shell model calculation for the respective states are 1.69 $\mu^2_N$, 1.53 $\mu^2_N$, and 1.30 $\mu^2_N$, from which we obtained a theoretical mean lifetime of 1.21 ps, 0.13 ps, and 0.045 ps. 
The mean lifetime measurement from the experimental 
and theoretical approach seems to agree. Further, the 13/2$^+$, 15/2$^+$ and 23/2$^+$
states in the band has B(M1) values of 0.28 $\mu^2_N$, 1.54 $\mu^2_N$, and 0.58 $\mu^2_N$ respectively. 
It is observed that the reduced transition probability B(M1) became strong at 15/2$^+$, reached a peak at 
17/2$^+$, and then started reducing gradually as spin increases. The highest B(M1) value here approaches the lower limit of 2 - 10 $\mu^2_N$ which has been observed for magnetic rotation bands in nuclei in higher mass region \cite{Clark}. 
Overall, the patterns seen in this band
till 23/2$^+$ is indicative of what is expected of the magnetic transition probabilities 
for the magnetic rotation sequence \cite{Clark}.

 The positive parity states in consideration can be viewed as part of collective 
 excitation because of the regularity observed in the $\gamma$ transitions linking the states. The presence of relatively
 strong $\mathrm{M1}$ transitions points to the possibility of a magnetic rotation band. 
 In magnetic transition bands, the magnetic transition probability, experimentally indicated by transition intensities is expected to decrease until the band terminates. 
 The g$_{9/2}$ orbit involved in the formation of positive parity states allows high spin to be generated 
 and also fulfills the neutron particle, proton hole coupling condition for 
 magnetic rotation as explained by Clark, and Macchiavelli \cite{Clark}. 
 The dominant configuration for the band of positive parity states is 
 $\pi$(f$^{-2}_{7/2}$p$^{1}_{3/2}$)$\otimes$$\nu$(p$^{2}_{3/2}$g$^{1}_{9/2}$). 
 The predicted occupancies of the states in this proposed magnetic rotation band according 
 to the shell model calculation can be seen from  Figure \ref{fig:Occupaancy} (a) 
 and highlight the very similar configuration for the whole band.
 
\subsection{$^{59}$Ni}

\begin{figure}
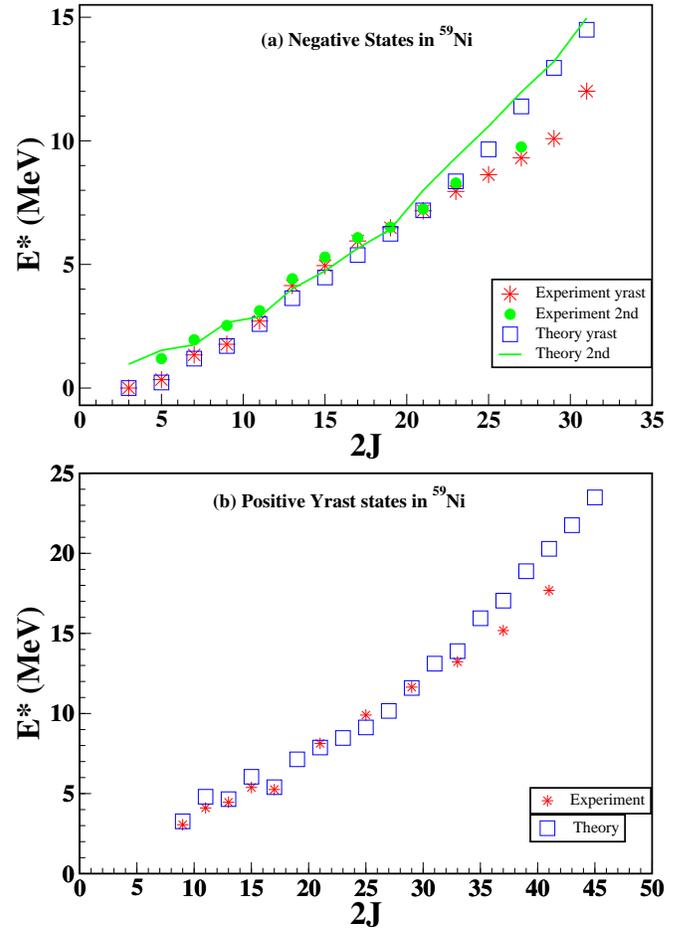

    \centering
    \includegraphics[clip,width=\columnwidth]{59Ni_NegYrast2nd_Line.eps}
    \includegraphics[clip,width=\columnwidth]{59Ni_yrast_Positive.eps}
    \caption{Comparison of theoretical shell model calculation with experimental results in $^{59}$Ni. 
    The red star symbol represents experimental values while the blue box symbol represents values 
    from the theoretical shell model calculation. In (a), the ``experiment 2nd" and the ``theory 2nd" are the second states (near y-rast, also called  yrare states), 
and are represented with green color; circles for experimental values, and line for the theoretical shell model calculation.}
    \label{fig:59Ni_ExpThe}
\end{figure}

$^{59}$Ni has a full 0f$_{7/2}$ orbital for both protons and neutrons, leaving 3 extra neutrons to fill the 
1p$_{3/2}$ orbit. With this configuration, the spin-parity of the ground state is 3/2$^-$ 
as a consequence of an unpaired neutron in the 1p$_{3/2}$ orbit. The first excited state with 
5/2$^-$ can be generated by the unpaired neutron moving from the 1p$_{3/2}$ orbital 
to the 0f$_{5/2}$ and is seen experimentally at 340 keV. The low-lying negative parity excited states have 
irregular energy transitions linking them that are suggestive of single-particle excitations within the 
$fp$ shell. Other nuclei in this A = 60 mass region have been found to exhibit similar single particle excitation 
for the low-lying states \cite{Anyageakaa, Soumik}. 
There is a good agreement between the shell model calculations for the negative parity states and experimental results up until the 23/2$^-$ as seen in 
Figure \ref{fig:59Ni_ExpThe} and Table~\ref{tab:Exp_Theo_Table}. 
Comparing the B(E2) values from our shell model calculation to the available evaluated
data for low lying states, it can be argued these states are non-collective in nature. 
Measured B(E2) values (in W.U) for the 7/2$^-$, 7/2$^-_2$, 9/2$^-$ and 11/2$^-$ states are 
given as 3.0(8), 12 (11), 10 (3) and 22 (5) respectively, 
while the theoretical calculations give 0.7, 16, 5.7 and 9.3 respectively.
While there is a good agreement between the shell model calculations and the experimental results 
for the low lying states, we see a pronounced variation from the 25/2$^-$ state and beyond. 
This discrepancy can be due to a different configuration for these high spin states 
likely involving more particles in the g$_{9/2}$ orbital. 
The excitation energy needed to generate these states in the collective model is likely 
lower than in the shell model picture.
 
The first excited positive parity state from the level scheme is a 9/2$^+$ state and can be easily 
generated by having a neutron in the g$_{9/2}$ orbital while all the other nucleons are paired in the 
$fp$ shell. The dominant configuration we, however, see from the theoretical prediction for the band 
of positive parity states is 
$\pi$(f$^{-1}_{7/2}$p$^{1}_{3/2}$)$\otimes$$\nu$(f$^{-1}_{7/2}$p$^{1}_{3/2}$f$^{1}_{5/2}$g$^{1}_{9/2}$)  
(see Figure \ref{fig:Occupaancy} (b) for the theoretical predictions). 
The shell model calculation for the yrast positive parity states aligns with the experimental 
results up till the 33/2$^+$ state. Further up, starting from the 37/2$^+$ state we again see a 
discrepancy between the shell model calculations and experimental results. 
This may also be due to a change in the configuration not captured by the shell model calculations as 
we move from the 33/2$^+$ level to 37/2$^+$. Such a  change of configuration was also observed in 
rotational bands in the study of $^{61}$Ni \cite{Soumik}, 
where the two configurations were illustrated by two fits to the rotational model.

Among the positive parity states, there is a sequence of $\mathrm{E2}$ transitions from 
9/2$^+$ to 41/2$^+$ which may be a pointer to rotational motion. Given the different configurations in different spin regions across this positive band, there seem to be different possible deformation configurations in this band. 
Since it is the rotation of a deformed nucleus that gives rise to regular bands observed in 
the spectrum of nuclei \cite{Clark}, it suggests $^{59}$Ni could be deformed at excitations involving the deformation driving g$_{9/2}$.
Rotation has been established in the four bands observed in the study of 
$^{59}$Ni by Yu {\it et al}. \cite{Yu}. 
The more populated bands 1 and 2 in the study correspond to the bands of positive parity states in our study. 
Cranked Nilsson-Strutinsky (CNS) calculations \cite{CNS} were performed in the study, and it indicated
substantial collectivity in the band structures of $^{59}$Ni. Similar deformed bands with a sequence of 
$\mathrm{E2}$ transitions were also established in the neutron-rich $^{61}$Ni by Bhattacharya {\it et al}. \cite{Soumik}.

The negative parity high spin states from 19/2$^-$ to 31/2$^-$ show a sequence of
$\mathrm{M1}$ transitions connecting the states. This could be an indication of magnetic rotation 
as was observed in $^{59}$Co. The configuration of these states could correspond to no valence nucleon 
in the g$_{9/2}$ orbit or 2 nucleons occupying the g$_{9/2}$ orbital. 
The dominant configuration of the 19/2$^-$ state according to our shell model calculation which cannot 
account for 2 particle excitation is given by 
$\pi$(f$^{-2}_{7/2}$ p$^{1}_{3/2}$)$\otimes$$\nu$(f$^{-1}_{7/2}$p$^{2}_{3/2}$ f$^{2}_{5/2}$). 
The  excitation from the f$_{7/2}$ orbital into the higher orbitals within the 
$fp$ shell could in principle generate proton and neutron angular momentum needed for magnetic 
rotation. Ref \cite{29Al_MagRot} indicated the possibility of magnetic rotation in 
the $sd$-shell region, where the highest available j orbital is d$_{5/2}$. 
This suggests there is a possibility for observing magnetic rotation without the 
inclusion of the g$_{9/2}$ orbital. The B(M1) for the states in consideration from 
the shell model calculations are all however
less than 0.1 $\mu^2_N$, which is small compared to the values for the M1 bands in
$^{59}$Co and $^{61}$Co (to be discussed). Ref. \cite{29Al_MagRot} also hinted that 
low B(M1) values will likely result from the unavailability of high j orbitals 
involved in the angular momentum coupling.

The implication of this result is that there might be two modes of excitation in 
this nucleus competing; 
one driven by the deformation of the nucleus, and the other, probably magnetic rotation due to the angular momentum 
coupling of the protons and neutrons. 
Though this is not new as it has been noted earlier in 
Refs \cite{Anyageakaa, 58Fe, 60Ni_Torres, Soumik} in their study of 
$^{61}$Co, $^{58}$Fe, $^{60}$Ni, and $^{61}$Ni, still it is 
interesting to see the extent of such a phenomenon with changing neutron numbers.

\subsection{$^{61}$Co}

\begin{figure}
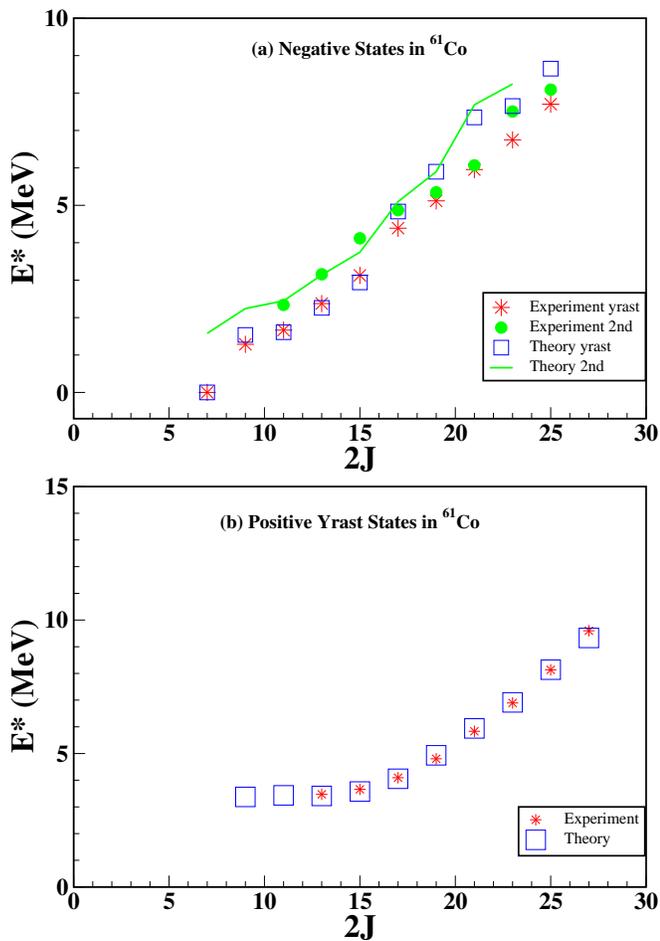

    \centering
    \includegraphics[clip,width=\columnwidth]{Negativeyrast_2nd_States61Co_Line.eps}
     \includegraphics[clip,width=\columnwidth]{Positive_yrastState61Co.eps}
    \caption{Comparison of theoretical shell model calculation with experimental results in $^{61}$Co. 
    The red star symbol represents experimental values while the blue box symbol represents values 
    from the theoretical shell model calculation. In (a), the ``experiment 2nd" and the ``theory 2nd" are the second states (near y-rast, also called  yrare states), 
and are represented with green color; circles for experimental values, and line for the theoretical shell model calculation.}
    \label{fig:61Co_ExpThe}
\end{figure}

$^{61}$Co with 27 protons and 34 neutrons has a ground state spin-parity of 7/2$^-$ just like $^{59}$Co. 
It can be seen as $^{59}$Co + 2n, so it is expected that excitations for the two isotopes have similarities. 
The first four excited states in the two nuclei are similar in energy levels. 
The first excited state, 9/2$^-$ is at 1.19 MeV and 1.29 
MeV in $^{59}$Co and $^{61}$Co respectively. The difference in the energies of the yrast states are 
95 keV at 9/2$^-$, 205 keV at 11/2$^-$, 219 keV at 13/2$^-$ and 211 keV at 15/2$^-$. Therefore, it can be 
expected that the same pattern of excitation will follow at low spins before the 
additional neutron pair is broken, and its excitation contributes to the spin of the nucleus. 
These considerable 
similarities in the low-lying spins are in agreement with the conclusions of Ref. \cite{59Co_61Co_1970} on the 
similarities between the energy levels of the low-lying states of $^{57}$Co, $^{59}$Co and $^{61}$Co.

The randomness of the transitions within the low-lying negative parity states in the level scheme suggests 
single particle excitation and seems to follow a $0p0h$ excitation pattern. 
Figure \ref{fig:61Co_ExpThe} and Table~\ref{tab:Exp_Theo_Table} shows the
comparison between the shell model calculation and the experimental results. 
Overall, a good agreement is observed, except for negative parity states from around 
21/2$^-$ upwards, where there seems to be a 
departure between the calculation and the experiment. 
It can be seen from the level scheme that the high-lying negative parity states are 
all linked by transitions to the 
13/2$^+$, 15/2$^+$ and 17/2$^+$ states which are $1p1h$ excitation and hence they could possibly represent 
$2p2h$ excitations, with two nucleons residing in the g$_{9/2}$ orbit. Configurations involving two 
g$_{9/2}$ neutrons were also suggested by Ref \cite{Anyageakaa} while discussing rotational bands in the 
high-lying negative parity states of $^{61}$Co however such calculations are not currently possible.

\begin{figure}
    \centering
    \includegraphics[clip,width=0.85\columnwidth]{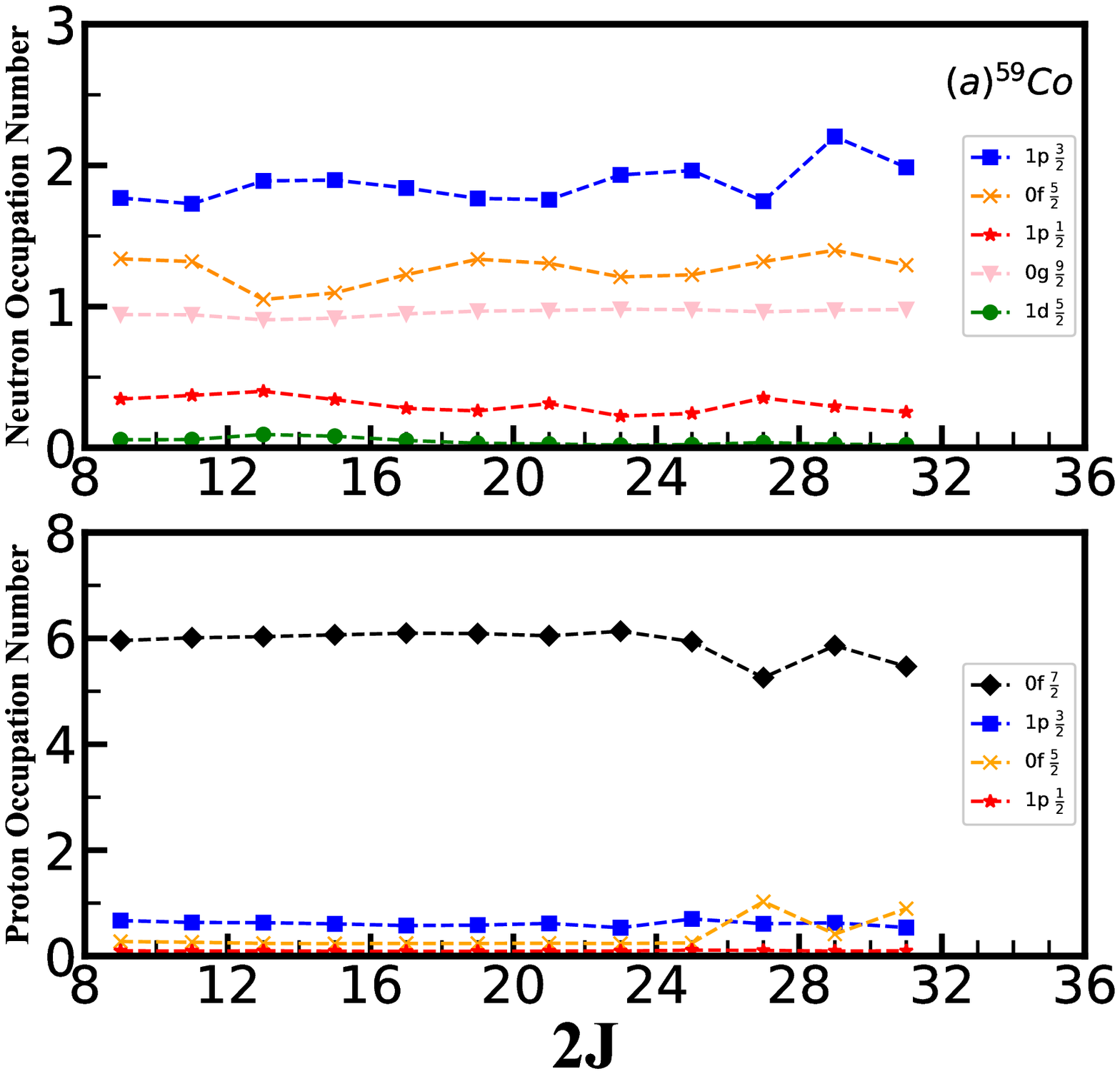}
    \includegraphics[clip,width=0.85\columnwidth]{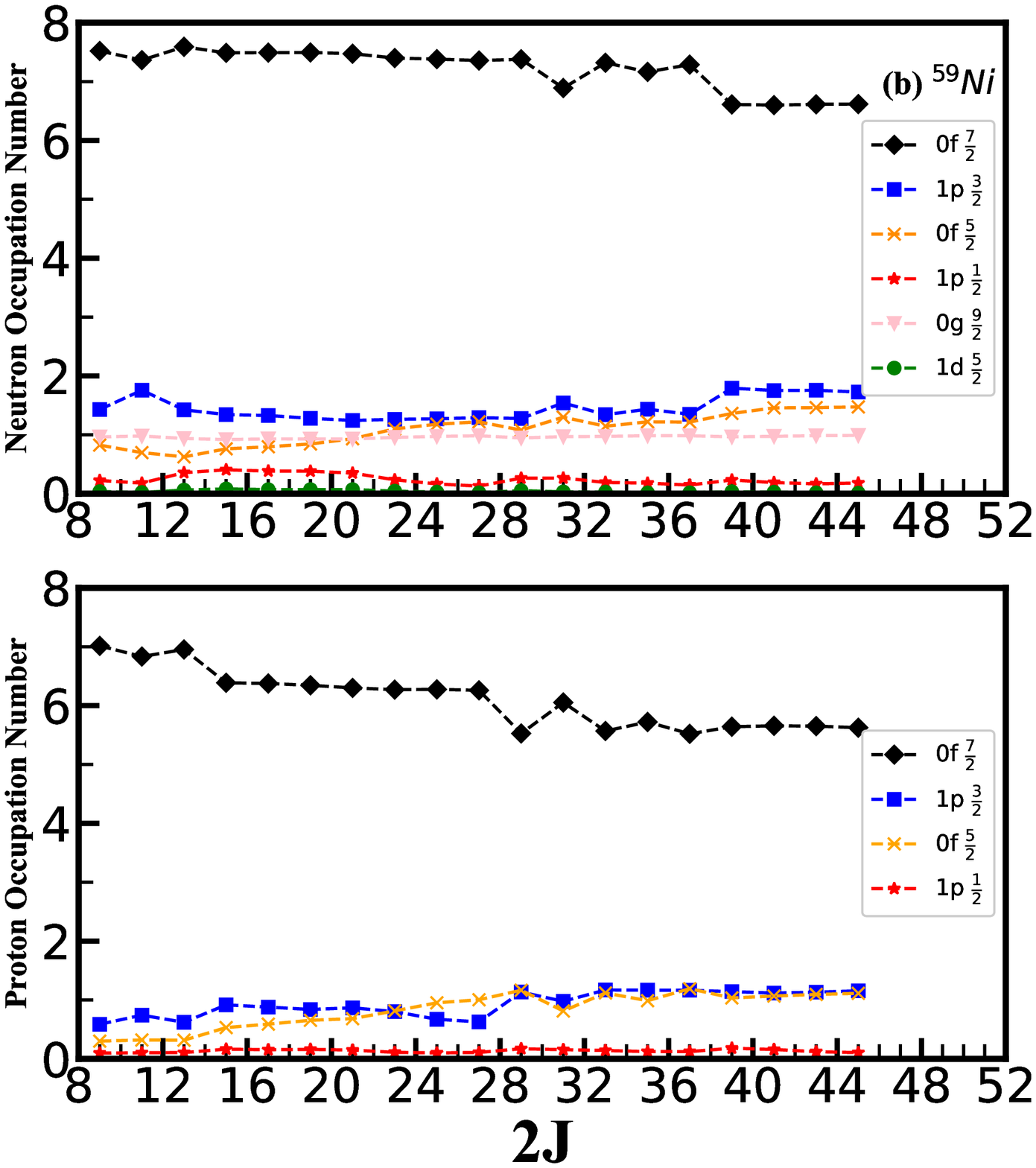}
    \includegraphics[clip,width=0.85\columnwidth]{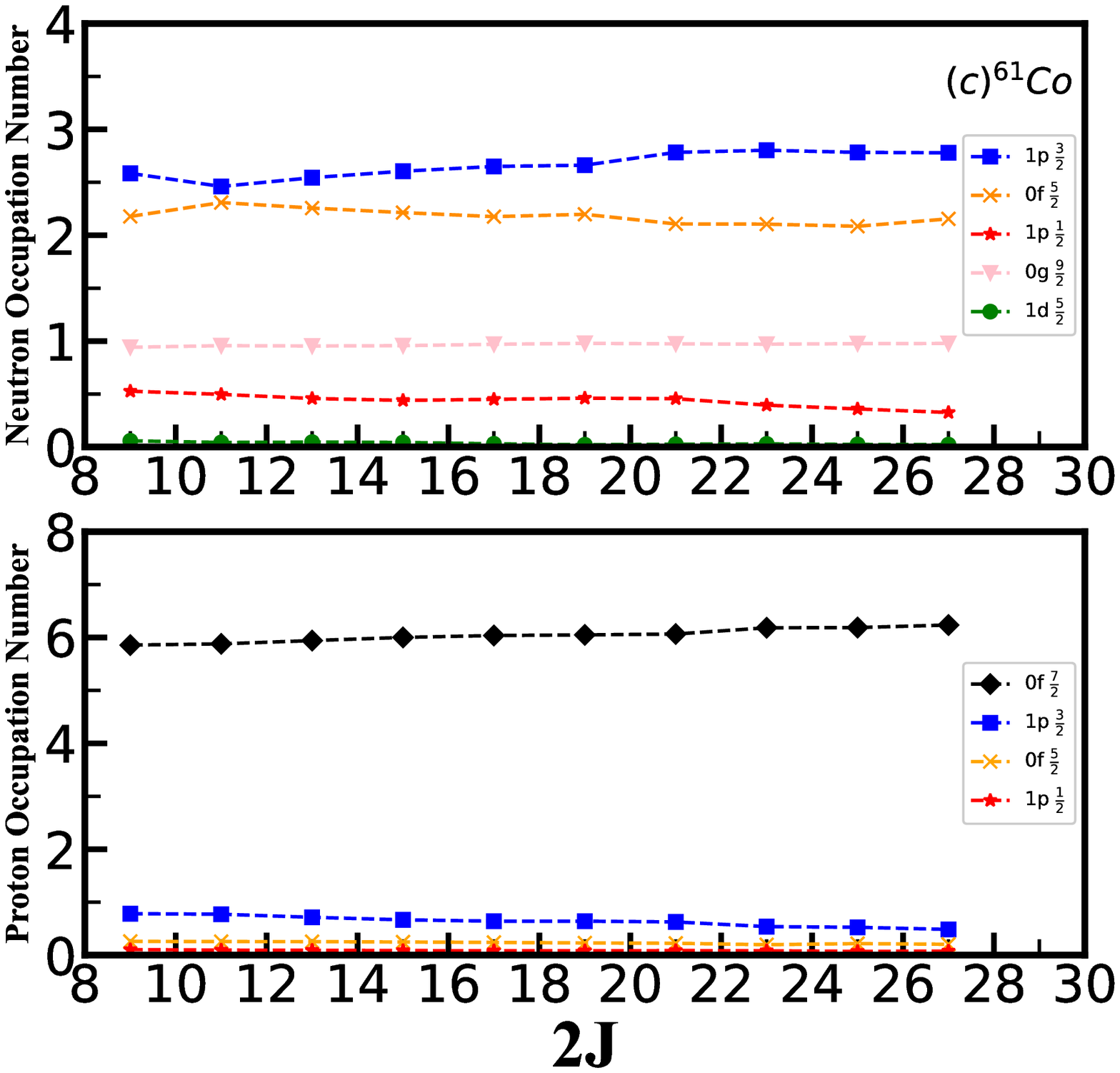}
     
    \caption{Occupancy of the protons and neutrons in the positive parity state in (a) $^{59}$Co 
(b) $^{59}$Ni and (c) $^{61}$Co as predicted by the shell model calculation. The orbitals 
are represented by different colors and shapes (blue box for p$_{3/2}$, orange X for f$_{5/2}$, 
red star for p$_{1/2}$, violet triangle for g$_{9/2}$, green circle for d$_{5/2}$ and black rhombus for f$_{7/2}$).}
    \label{fig:Occupaancy}
\end{figure}

The positive parity states assigned in this study of $^{61}$Co are a natural extension of what we observed 
in $^{59}$Co. In Ref.~\cite{Anyageakaa} these were not proposed as 
positive parity states as linear polarization measurement was not performed in that study. 
They did though observe a difference in the excitation pattern above 4 MeV, (which corresponds to 17/2$^+$) 
and therefore suggested that the description of higher-lying states in $^{61}$Co should be carried out in an 
expanded model space beyond the $fp$ shell \cite{Anyageakaa}, which has been done in this current study to 
predict the positive parity states. From Figure \ref{fig:61Co_ExpThe}, we see a good 
agreement between the shell 
model calculations for the positive parity states and the experimental results. 
The occupancies associated with the positive parity band are displayed in 
Figure \ref{fig:Occupaancy} (c).
The regularity observed in this band of positive parity states which contains a series of $\mathrm{M1}$ 
transitions is again an indication of magnetic rotation. 
Experimental value for the ``apparent" lifetime of the
17/2$^+$ state has been measured by Ref \cite{61_63Co} and Ref \cite{A=61} as 1.1(0.3) ps.
The B(M1) from the shell model calculation for the strongest transition from the 17/2$^+$ state is 1.0 $\mu^2_N$,
from which we obtained a theoretical partial mean lifetime of 0.69 ps for the state. Converting this theoretical partial mean lifetime into total mean lifetime using the branching 
ratio from Table \ref{tab:61Co} gives a value of 0.57 ps. The ``apparent" lifetime from Ref \cite{61_63Co} represents an upper limit for the experimental mean lifetime of the state. 
The theoretical B(M1) values were obtained for other states, 15/2$^+$, 19/2$^+$, 21/2$^+$, 23/2$^+$ in the band and are  1.17 
$\mu^2_N$, 0.81 $\mu^2_N$, 0.22 $\mu^2_N$, and 0.5 $\mu^2_N$ respectively. 
The values shows an approximate pattern expected of the magnetic transition 
probabilities (reduction) with spin increase as explained by Ref \cite{Clark}. 

The dominant configuration for the band of positive parity states according to shell model calculation is 
$\pi$(f$^{-2}_{7/2}$p$^{1}_{3/2}$)$\otimes$$\nu$(p$^{-1}_{3/2}$f$^{2}_{5/2}$g$^{1}_{9/2}$) 
(Figure \ref{fig:Occupaancy} (c)). 
The other bands of high-lying negative parity states also have $\mathrm{M1}$ 
transitions and could be candidates for magnetic rotation too. 
We also observe the established band of E2 transitions indicative of rotation due to deformation, 
but it terminates quickly in this study. Just as was discussed for $^{59}$Ni, 
different modes of excitations are observed in the case of $^{61}$Co. 
Ref \cite{Anyageakaa} concluded that in $^{61}$Co, quadrupole collectivity 
associated with a prolate shape 
competes for yrast status with the magnetic rotation of a nearly spherical system. 
Our results in general also support the notion that there is a competition between 
the magnetic rotation and the rotation due to deformation in this nucleus.

\section{Summary}

In the current work, we have studied the structure of three nuclei with $A \approx 60$; 
namely $^{59}$Co, $^{59}$Ni and $^{61}$Co using a $^{14}$C beam on $^{48}$Ti and $^{50}$Ti targets.
The level scheme of $^{59}$Co now includes positive parity states which has been extended to 
31/2$^+$ at around 11 MeV. The positive parity states in the $^{59}$Ni level scheme are now 
extended to 41/2$^+$ at excitation energy 17.7 MeV. The negative parity states have also 
been extended to 31/2$^-$ at excitation energy of 12 MeV. 
There were no new transitions observed for $^{61}$Co, but one of the major bands has been 
reassigned as positive parity states by reason of this study.

It can be concluded from this study that the low-lying negative parity states in the three nuclei 
represent single particle excitations of the nucleons limited to the $fp$ shell. 
The positive parity states and possibly the high-lying negative parity states in
these nuclei are highly influenced by the presence of nucleons in the g$_{9/2}$ orbit. The 
presence of either strong $E2$ or $M1$ transitions observed in these nuclei suggests a 
form of collective excitation. We observe that collective excitation in
$^{59}$Co is dominated by magnetic rotation which can be explained by the shears mechanism. 
However, the collective excitations in $^{59}$Ni and $^{61}$Co are both in the form of
magnetic rotation (though only tentative indications in $^{59}$Ni) and rotations
due to deformation. 
This implies that the excitation of neutrons into the g$_{9/2}$ orbit creates deformation in 
$^{59}$Ni and $^{61}$Co which in turn brings about a rotation that competes with the magnetic 
rotation occurring with little to no deformation. 
The three nuclei studied highlight the competition between single particle excitations and collective 
excitations in the excited states observed. The large scale shell model calculations presented here 
reproduced the excited states for the most part. However, the departures at 
the very high spins clearly point to the need for refinements of calculation to include more than 
one nucleon moving to the $g_{9/2}$ orbit.

\section{Acknowledgement}

This work was supported by the U.S. National Science Foundation under grant number Phy-2012522 (FSU) 
and the U.S. Department of Energy, office of Science, under award number DE-AC05-00OR22725 (ORNL).  
Noritaka Shimizu and Yutaka Utsuno acknowledge the support by \'{A}\'{A}Program for Promoting Researches 
on the Supercomputer Fugaku \copyright 
 (JPMXP1020200105) and Multidisciplinary Cooperative Research Program, Tsukuba-CCS (wo22i002).

\providecommand{\noopsort}[1]{}\providecommand{\singleletter}[1]{#1}%
%

\newpage
\begin{appendices}
\begin{longtable*}[c]{|c c c c c c c c c|}
\caption{This table show the details of the $\gamma$ transitions observed in the present work on $^{59}$Co. The DCO ratio ($R_{DCO}$) was calculated for quadrupole transitions except for those with the D symbol which are from dipole transitions. The relative intensities, I$_\gamma$ less than 6\% are given to 1 decimal place. \label{tab:59Co}}\\
 
 \hline
 \multicolumn{9}{|c|}{Begin of Table}\\
 \hline
 E$_i$ (keV) & E$_f$ (keV) & E$_\gamma$ (keV)  & I$_\gamma$ (\%) & J$^{\pi}_i$ & J$^{\pi}_f$ & $R_{DCO}$ & Asymmetry & Multipolarity \\
 \hline
 \endfirsthead
 
 \hline
 \multicolumn{9}{|c|}{Continuation of Table \ref{tab:59Co}}\\
 \hline
  E$_i$ (keV) & E$_f$ (keV) & E$_\gamma$ (keV)  & I$_\gamma$ (\%) & J$^{\pi}_i$ & J$^{\pi}_f$ & $R_{DCO}$ & Asymmetry & Multipolarity \\
 \hline
 \endhead

 \hline
 \endfoot
 
\hline
 \multicolumn{9}{| c |}{End of Table}\\
 \hline
 \endlastfoot
        1190.7 (5) & 0 & 1190.7 (5) & 100 (11) & 9/2$^-$ & 7/2$^-$ & 0.91 (6)$^D$ & -0.014 (10) & M1\\
        1460.4 (5) & 0 & 1460.4 (5) & 100 (9) & 11/2$^-$ & 7/2$^-$ & 1.63 (9)$^D$ & 0.013 (13) & E2 \\
        & 1190.7 & 269.2 (5) & 13 (1) & 11/2$^-$ & 9/2$^-$ & 1.26 (9)$^D$ & & D\\
        2154.9 (7) & 1460.4 & 694.5 (5) & 74 (7) & 13/2$^-$ & 11/2$^-$ & 0.63 (5) & -0.0077 (11) & M1\\
        2183.7 (8) & 1190.7 & 993.0 (6) & 77 (11) & 11/2$^-$ & 9/2$^-$ & 1.03 (7)$^D$ && D\\
        & 0 & 2183.7 (8) & 3.2 (3) & 11/2$^-$ & 7/2$^-$ && &\\
        2915.4(10) & 2154.9	& 760.5 (7) & 21 (2) & 15/2$^-$ & 13/2$^-$ & 0.68 (5) & -0.0005 (9) & M1/E2 \\
        3083.4 (8) & 1190.7 & 1892.7 (6) & 5.2 (5) & 11/2$^-$ & 9/2$^-$ & 1.04 (7)$^D$ & -0.057 (18) & M1\\
        3224.9 (9) & 2183.7 & 1041.2 (5) & 5.0 (3) & 13/2$^{(-)}$ & 11/2$^-$ & 1.00 (14)$^D$ & & D\\
        & 1460.4 & 1764.9 (9) & 6 (1) & 13/2$^{(-)}$ & 11/2$^-$ & 1.07 (11)$^D$&& D\\
        3627.5(11) & 2915.4 & 712.1 (5) & 16 (2) & 17/2$^{(-)}$ & 15/2$^-$ & 0.62 (5) && D\\
        3738.4(9) & 2154.9 & 1583.5(5) & 11 (1) & 15/2$^-$ & 13/2$^-$ & 0.71 (5) & -0.0016 (10) & M1\\
        3844.3(10) & 1190.7 & 2653.6 (6) & 12 (1) & 11/2$^+$ & 9/2$^-$ & 0.81 (6)$^D$ & 0.075 (21) & E1 \\
        & 1460.4 & 2383.9 (9) & 7 (1) & 11/2$^+$ & 11/2$^-$ & 0.98 (7) & 0.075 (14) & E2, $\Delta$J = 0 \\
        & 2154.9 & 1690.2 (8) & 4.9 (3) & 11/2$^+$ & 13/2$^-$ & 0.79 (6)$^D$ & & D\\
        4178.7(11) & 2183.7 & 1995.3 (5) & 33 (2) & 13/2$^+$ & 11/2$^-$ & 0.97 (7)$^D$ & 0.066 (13) & E1 \\
        & 3844.3 & 334.4 (5) & 14 (1) & 13/2$^+$ & 11/2$^+$ & 0.97 (7)$^D$ & & D\\
        & 1460.4 & 2719.4 (5) & 10 (1) & 13/2$^+$ & 11/2$^-$ & 0.54 (5) & 0.083 (22)  & E1\\
        & 3083.4 & 1095.1 (5) & 5.0 (4) & 13/2$^+$ & 11/2$^-$ & 0.62 (5) & & D\\
        & 2154.9 & 2025.3 (5) & 4.5 (3) & 13/2$^+$ & 13/2$^-$ &&& $\Delta$J = 0  \\
        4414.6(11) & 4178.7 & 235.9 (5) & 41 (3) & 15/2$^+$ & 13/2$^+$ & 1.14 (3)$^D$ & & D\\
        & 2154.9 & 2260.6 (5) & 17 (1) & 15/2$^+$ & 13/2$^-$ & 0.48 (5) & 0.055 (13) & E1 \\
        & 3224.9 & 1190.3 (6) & 4.8 (4) & 15/2$^+$ & 13/2$^-$ & 0.91 (6)$^D$ & & D\\
        4489.6(9) & 1460.4 & 3029.2 (8) & 3.5 (2) & 15/2$^{(-)}$ & 11/2$^-$ & 1.25 (18)&& Q\\
        4717.5(12) & 4414.6 & 302.9 (5) & 55 (7) & 17/2$^+$ & 15/2$^+$ & 0.68 (5) & & D\\
        4785.2(13) & 3738.4 & 1046.8 (9) & 3.4 (3) & 17/2$^{(-)}$ & 15/2$^-$ & 1.23 (9)$^D$ && D\\
        4799.9(12) & 3627.5 & 1172.4 (5) & 13 (6) & 19/2$^{(-)}$ & 17/2$^{(-)}$ & 0.67 (5) & & D\\
        5370.6(13) & 4717.5 & 653.1 (5) & 53 (5) & 19/2$^+$ & 17/2$^+$ & 0.58 (4) & -0.017 (12) & M1\\
        5444.8(11) & 2154.9	& 3289.9 (9) & 3.4 (3) & 17/2$^{(-)}$ & 13/2$^-$ & 1.19 (13) && Q\\
        5574.8(11) & 2154.9 & 3419.9 (9) & 3.4 (3) & 15/2$^{(-)}$ & 13/2$^-$ & 0.48 (5) && D\\
        6364.6(14) & 5370.6 & 994.0 (6) & 25 (4) & 21/2$^+$ & 19/2$^+$ & 0.50 (5) && D\\
        6422.4(13) & 4799.9 & 1622.5 (5) & 2.9 (3) & 19/2$^{(-)}$ & 19/2$^{(-)}$ & 1.14  (8) && $\Delta$J = 0 \\
        & 3627.5 & 2794.0 (7) & 2.8 (2) & 19/2$^{(-)}$ & 17/2 $^{(-)}$ & 0.70 (8) && D\\
       6878.7(14) & 5370.6 & 1508.9 (5) & 5.1 (5) & 21/2$^{(-)}$ & 19/2$^+$ & 0.50 (5) && D\\
        & 4799.9 & 2078.8 (5) & 10 (1) & 21/2$^{(-)}$ & 19/2$^{(-)}$ & 0.54 (5) & & D\\
        & 6422.4 & 457.1 (5) & 4.1 (4) & 21/2$^{(-)}$ & 19/2$^{(-)}$ & 0.68 (5) & -0.0035 (10) & M1\\
        7459.7(15) & 6364.6 & 1095.1 (5) & 9 (1) & 23/2$^+$ & 21/2$^+$ & 0.62 (5) & -0.020 (13) & M1\\
        7842.7(15) & 6364.6 & 1478.1 (5) & 10 (1) & 23/2$^{(+)}$ & 21/2$^+$ & 0.71 (5) && D\\
        7903.9(15) & 6364.6 & 1539.3 (5) & 5.0 (4) & 23/2$^{(-)}$ & 21/2$^+$ & 1.16 (8)$^D$ & 0.013 (12) & E1\\
        & 6878.7 & 1025.6 (6) & 5.0 (4) & 23/2$^{(-)}$ & 21/2$^{(-)}$ & 0.64 (5) & -0.0080 (11) & M1\\
        8118.5(16) & 6878.7 & 1239.8(8) & 3.7 (3) & 23/2$^{(-)}$ & 21/2$^{(-)}$ & 0.73 (5)& & D\\
        8868.1 (16) & 7903.9 & 964.2 (6) & 3.9 (3) & 25/2$^{(-)}$ & 23/2$^{(-)}$ & 0.66 (6) & & D\\
        8898.2(16) & 7459.7 & 1437.6 (6) & 5.0 (5) & 25/2$^{(+)}$ & 23/2$^+$ & 0.41 (5) && D\\
        9105.4(17) & 7459.7 & 1644.8 (9) & 3.5 (3) & 25/2$^{(+)}$ & 23/2$^+$ & 1.14 (8)$^D$ && D\\
        9343.8(15) & 6364.6 & 2979.2 (5) & 3.6 (3) & 25/2$^{(+)}$ & 21/2$^+$ & 2.08 (23)$^D$ & 0.027 (15) & E2\\
        9552.9(16) & 7903.9 & 1649.0 (5) & 3.9 (2) & 27/2$^{(-)}$ & 23/2$^{(-)}$ & 1.06 (11)&& Q\\
        9815.1(16) & 7842.7 & 1971.5 (5) & 4.1 (4) & 27/2$^{(+)}$ & 23/2$^{(+)}$ & 0.94 (7) & 0.21 (4) & E2\\
        10073.2(16) & 9343.8 & 729.4 (5) & 3.2 (5) & 27/2$^{(+)}$ & 25/2$^{(+)}$ & 0.68 (5) & & D\\
        10091.0(16) & 7459.7 & 2630.4 (6) & 2.9 (2) & 27/2$^{(+)}$ & 23/2$^+$ & 1.13 (13) & 0.12 (5) & E2\\
        10888.7(17) & 10091.0 & 795.7 (7) & 2.7 (4) & 31/2$^{(+)}$ & 27/2$^{(+)}$ & 1.25 (8) && Q\\
        11140.1(18) & 9815.1 & 1325.0 (8) & 3.5 (5) & 31/2$^{(+)}$ & 27/2$^{(+)}$ & 1.64 (12)$^D$&& Q\\

   \end{longtable*}

\begin{longtable*}[c]{|c c c c c c c c c|}
\caption{This table show the details of the $\gamma$ transitions observed in the present work on $^{59}$Ni. The $R_{DCO}$ was 
calculated for quadrupole transitions except for those with the D symbol which are from dipole transitions. The relative intensities, 
I$_\gamma$ less than 6\% are given to 1 decimal place. \label{tab:59Ni}}\\
 
 \hline
 \multicolumn{9}{|c|}{Begin of Table}\\
 \hline
 E$_i$ (keV) & E$_f$ (keV) & E$_\gamma$ (keV)  & I$_\gamma$ (\%) & J$^{\pi}_i$ & J$^{\pi}_f$ & $R_{DCO}$ & Asymmetry & Multipolarity \\
 \hline
 \endfirsthead
 
 \hline
 \multicolumn{9}{|c|}{Continuation of Table \ref{tab:59Ni}}\\
 \hline
  E$_i$ (keV) & E$_f$ (keV) & E$_\gamma$ (keV)  & I$_\gamma$ (\%) & J$^{\pi}_i$ & J$^{\pi}_f$ & $R_{DCO}$ & Asymmetry & Multipolarity \\
 \hline
 \endhead

 \hline
 \endfoot
 
\hline
 \multicolumn{9}{| c |}{End of Table}\\
 \hline
 \endlastfoot
        339.5(5) &	0 &	339.5(5) & 100 (10) & 5/2$^-$ & 3/2$^-$ & 0.74 (5) && D\\
        1189.0(5)	 & 0 & 1189.0(5)* & 53 (16) & 5/2$^-$ & 3/2$^-$ & 0.52 (6) && D\\
        1339.0(6)	& 339.5 & 998.0(5)* & 61 (15) & 7/2$^-$ & 5/2$^-$ & 0.96 (6)$^D$ && D\\
        & 0	& 1339.0(6) & 39 (7) & 7/2$^-$ & 3/2$^-$ & 1.67 (16)$^D$ & 0.053 (11) & E2\\
        1768.5(7) & 339.5 & 1429.0(5) & 63 (9) & 9/2$^-$ & 5/2$^-$ & 0.93 (13) & 0.093 (13) & E2\\
        &1339.0 &	429.1(8) & 18 (2) & 9/2$^-$ & 7/2$^-$ & 0.87 (9)$^D$&& D\\
        1949.7(7) &	0 &	1949.7(7) & 16 (3) & 7/2$^-$ & 3/2$^-$ & 0.94 (7) & 0.079 (13)& E2\\
        &339.5 & 1610.4(5) & 9 (4) & 7/2$^-$ & 5/2$^-$ & 0.58 (4) && D\\
        &1189.0 &	759.0(5) & 6 (1) & 7/2$^-$ & 5/2$^-$ & 0.56 (7)&& D\\
        &1339.0 & 610.0 (5) & 4.4 (2) & 7/2$^-$ & 7/2$^-$ & 1.79 (17)$^D$&&$\Delta$J = 0 \\
        2530.0(8)	&1339.0 & 1191.0(5) & 6 (1) & 9/2$^-$ & 7/2$^-$ & 0.90 (9)$^D$ && D\\
        & 339.5 &	2193.0(6) & 2.1 (3) & 9/2$^-$ & 5/2$^-$ & 1.74 (17)$^D$&& Q\\
        2707.0(7) & 1339.0 & 1368.0(5) & 58 (4) & 11/2$^-$ & 7/2$^-$ & 1.50 (14)$^D$& 0.13 (1) & E2\\
        &1768.5 &	938.3(5) & 15 (1) & 11/2$^-$ & 9/2$^-$ &  0.42 (3) && D\\
        3056.6(7) & 1949.7 & 1105.9(8) &  26 (4) & 9/2$^+$ & 7/2$^-$ & 0.57 (4) & 0.086 (12) & E1\\
        & 1339.0 & 1717.6(5) & 9 (2) & 9/2$^+$ & 7/2$^-$ & 0.59 (4) & 0.040 (12) & E1\\
        3126.5(9) & 1768.5 &	1358.0(6) & 3.3 (3) & 11/2$^-$ & 9/2$^-$ & 0.88 (11)$^D$&& D\\
        & 339.5 & 2783.0(7) & 2.1 (4) & 11/2$^-$ & 5/2$^-$ &&&\\
        3378.2(9) & 1768.5 & 1609.7(5) & 24 (1) & 11/2$^-$ & 9/2$^-$ & 0.66 (7) && D\\
        &2707.0 & 674.0(5) & 22 (3) & 11/2$^-$ & 11/2$^-$ & 0.89 (9) &&$\Delta$J = 0\\
        3561.8(8) &1768.5 &	1793.3 (5) & 24 (4) & 11/2$^-$ & 9/2$^-$ & 0.71 (5) & -0.0040 (11)& M1\\
        & 2707.0	& 854.6(8) & 22 (3) & 11/2$^-$ & 11/2$^-$ &&& $\Delta$J = 0 \\
        &2530.0	& 1029.0(5) & 6 (3) & 11/2$^-$ & 9/2$^-$ & 1.09 (11)$^D$&& D\\
        & 3126.5 &	433.6(5) & 5.0 (6) & 11/2$^-$ & 11/2$^-$ & 1.27 (13) && $\Delta$J = 0\\
        &1949.7 &	1612.0(5) & 5.0 (5) & 11/2$^-$ & 7/2$^-$ &&&\\
         & 1339.0 & 2223.2(6) & 3.8 (3) & 11/2$^-$ & 7/2$^-$ & 1.07 (8) && Q\\
        4104.7(9) & 1768.5	& 2336.2(5) & 6 (1) & 11/2$^+$ & 9/2$^-$ & 0.76 (5)& 0.062 (19) & E1\\
        4143.8(9) & 3378.2 & 764.9(5) & 31 (2) & 13/2$^-$ & 11/2$^-$ & 0.99 (8)$^D$ & -0.0005 (9) & M1\\
        & 3561.8 &	582.0(5) & 27 (2) & 13/2$^-$ & 11/2$^-$ & 0.90 (9)$^D$ & -0.033 (9)& M1\\
        &	2707.0 &	1435.0(5) & 2.2 (3) & 13/2$^-$ & 11/2$^-$ &&&\\
        4419.0(10) &	1768.5 & 2650.5(8) & 6 (1) & 13/2$^-$ & 9/2$^-$ & 1.15 (10) & 0.058 (18) & E2\\
        & 3561.8 &	858.0(8) & 2.7 (2) & 13/2$^-$ & 11/2$^-$ &&&\\
        4457.8(9) & 3056.6 &	1401.5(5) & 25 (2) & 13/2$^+$ & 9/2$^+$ & 0.82 (6) &0.069 (12) & E2\\
        &	2707.0	&1750.8(5) & 24 (2) & 13/2$^+$ & 11/2$^-$ & 0.44 (6) & 0.054 (12) & E1 \\
        4727.2(9) & 1768.5	& 2958.7(5) & 3.1 (3) & 11/2$^+$ & 9/2$^-$ & 0.56 (8) & 0.030 (14)& E1\\
        4910.7(10) & 1768.5 & 3142.2(7) & 4.4 (3) & 11/2$^+$ & 9/2$^-$ & 0.64 (9) & 0.070 (24) & E1\\
        4950.5(10) & 4143.8	& 806.7(5) & 29 (4) & 15/2$^-$ & 13/2$^-$ & 0.87 (9)$^D$ & -0.055 (12)& M1\\
        &	3378.2&	1571.0(6) & 3.1 (2) & 15/2$^-$ & 11/2$^-$ &&&\\
        5100.0(9) & 3561.8 & 1539.0(7) & 3.7 (2) & 13/2$^{(-)}$ & 11/2$^-$ & 1.27 (13)$^D$&& D\\
        & 2707.0 & 2393.0(6) & 2.3 (2) & 13/2$^-$ & 11/2$^-$ &&& \\
        5254.6(10) & 4457.8	& 796.8(5) & 32 (2) & 17/2$^+$ & 13/2$^+$ & 1.00 (7) & 0.051 (25) & E2\\
        &4950.5 & 303.0(5) & 2.9 (1) & 17/2$^+$ & 15/2$^-$ & 1.08 (11)$^D$ && D\\
        5295.0(11) &	4143.8 	& 1151.2(7) & 4.5 (3) & 15/2$^-$ & 13/2$^-$ & 0.78 (8)$^D$ & -0.025 (14) & M1\\
        5383.2(10) &4143.8 & 1239.4(6) & 10 (1) & 15/2$^+$ & 13/2$^-$ & 1.29 (13)$^D$ & 0.10 (1) & E1\\
        & 4104.7 & 1279.3(5) & 4.4 (4) & 15/2$^+$ & 11/2$^+$ & 1.08 (8) && Q\\
        5947.7(11) & 4950.5 & 997.2(5) & 16 (4) & 17/2$^{(-)}$ & 15/2$^-$ & 0.94 (7)$^D$&& D\\
        5991.8(11) & 4457.8 & 1534.0(6) & 5.0 (5) & 17/2$^+$ & 13/2$^+$ &&&\\
        6079.3(10) & 4143.8 & 1935.5(5) & 14 (2) & 17/2$^-$ & 13/2$^-$ & 0.98 (7) & 0.069 (20) & E2\\
        6483.8(12) & 5947.7 & 536.1(6) & 5.0 (4) & 19/2$^{(-)}$ & 17/2$^{(-)}$ &  0.57 (4) & -0.0017 (99)& M1\\
        6505.6(12) & 6079.3 & 426.6(5) & 11 (2) &  19/2$^-$ & 17/2$^-$ & 0.70 (5) && D\\
        & 4950.5 & 1555.1(7) & 10 (1) & 19/2$^-$ & 15/2$^-$ & 1.32 (9) & 0.037 (11) & E2\\
        6749.6(10) & 5100.0 &	1649.6(5) & 3.1 (2) & 15/2$^{(-)}$ & 13/2$^{(-)}$ & 0.87 (9)$^D$&& D\\
        7167.5(13) & 6505.6 & 661.9(5) & 17 (2) & 21/2$^{(-)}$ & 19/2$^-$ & 0.71 (5) && D\\
        & 5254.6 & 1912(9) & 2.1 (1) & 21/2$^-$ & 17/2$^+$ &&&\\
        7232.7(10) & 6483.8 & 748.9(6) & 4.3 (4) & 21/2$^{(-)}$ & 19/2$^{(-)}$ & 0.7 (1)&& D\\
        7393.7(13) &	5254.6 &	2139.1(8) & 5.5 (4) & 19/2$^-$ & 17/2$^+$ & 0.75 (6) & 0.021 (11) & E1\\
        7636.8(13) &	6505.6 &	1131.2(5) & 3.0 (2) & 21/2$^{(-)}$ & 19/2$^-$ & 0.87 (10)$^D$ && D\\
        7954.2(14) &	7167.5 &	786.7(5) & 11 (1) & 23/2$^{(-)}$ & 21/2$^{(-)}$ & 0.48 (5) & -0.024 (11) & M1\\
        8132.8(13) &	5254.6 &	2878.2(7) & 14 (1) & 21/2$^+$ & 17/2$^+$ & 1.16 (10)& 0.052 (13) & E2\\
        8290.7(14) &	7393.7 &	897.0(5) & 4.9 (2) & 23/2$^-$ & 19/2$^-$ & 0.95 (7) & 0.011 (11)& E2\\
        8632.3(15) &	7954.2 &	678.1(5) & 10 (3) & 25/2$^{(-)}$ & 23/2$^{(-)}$ & 0.58 (4) && D\\
        9166.5(15) &	8290.7 &	875.8(6) & 4.2 (4) & 27/2 & 23/2$^-$ & 1.12 (7) && Q\\
        9309.7(16) &	8632.3 &	677.4(5) & 10 (2) & 27/2$^{(-)}$ & 25/2$^{(-)}$ & 0.58 (4) && D\\
        9753.4(16) &	8632.3 &	1121.1(5) & 2.1 (1) & 27/2$^{(-)}$ & 25/2$^{(-)}$ & 1.14 (14)$^D$& -0.12 (2) & M1\\
        9901.5(14) &	8132.8 &	1768.7(5) & 7 (1) & 25/2$^+$ & 21/2$^+$ & 1.08 (8) & 0.084 (12) & E2\\
        10089.1(17) &	9309.7 &	779.4(5) & 5.0 (4) & 29/2$^{(-)}$ & 27/2$^{(-)}$ & 0.63 (7) && D\\
        10421.8(14) &	8132.8	& 2289.0(5) & 4.7 (2) & 25/2$^+$ & 21/2$^+$ & 1.26 (18) & 0.072 (20) & E2\\
        11645.0(15) &	9901.5 &	1747.0(5) & 3.8 (1) & 29/2$^+$ & 25/2$^+$ & 1.61 (23)$^D$ && Q\\
        & 10421.8 &	1223.2 (5) & 3.7 (1) & 29/2$^+$ & 25/2$^+$ & 1.58 (16)$^D$& 0.047 (11) & E2\\
        11911.8(16) &	9901.5 &	2010.3(8) & 3.7 (4) & 29/2$^+$ & 25/2$^+$ & 1.00 (7) & 0.11 (2)& E2\\
        12007.5(18)	& 10089.1	& 1918.4(8) & 2.5 (5) & 31/2$^{(-)}$ & 29/2$^{(-)}$ & 0.96 (14)$^D$&& D\\
        13229.0(17) &	11645.0 &	1584.0(7) & 5.2 (3) & 33/2$^{(+)}$ & 29/2$^+$ & 1.77 (22)$^D$&& Q\\
        14284.1(17) &	11911.8 &	2372.3(6) & 3.3 (3) & 33/2$^{(+)}$ & 29/2$^+$ & 1.08 (10)&& Q\\
        15178.7(18) & 13229.0 & 1949.7(7) & 4.0 (5) & 37/2$^{(+)}$ & 33/2$^{(+)}$ & 0.94 (7) & 0.079 (13)& E2\\
        16467.6(19) & 14284.1 & 2183.5(9) & 1.6 (7) & (37/2$^+$) & 33/2$^+$ &&&\\
        17684.6(19)	& 15178.7 &	2505.9(5) & 3.1 (3) & 41/2$^{(+)}$ & 37/2$^{(+)}$ &&&\\
  \end{longtable*}
*The intensity measurement of the 1189.0- and 998.0-keV transition is complicated by nearby contaminant transitions. \\\\\\\\

\begin{longtable*}[c]{|c c c c c c c c c|}
\caption{This table show the details of the $\gamma$ transitions observed in the present work on $^{61}$Co. 
The $R_{DCO}$ was calculated for quadrupole transitions except for those with the D symbol which are from dipole transitions. The relative intensities, I$_\gamma$ less than 6\% are given to 1 decimal place.\label{tab:61Co}}\\
 
 \hline
 \multicolumn{9}{|c|}{Begin of Table}\\
 \hline
 E$_i$ (keV) & E$_f$ (keV) & E$_\gamma$ (keV)  & I$_\gamma$ (\%) & J$^{\pi}_i$ & J$^{\pi}_f$ & $R_{DCO}$ & Asymmetry & Multipolarity \\
 \hline
 \endfirsthead
 
 \hline
 \multicolumn{9}{|c|}{Continuation of Table \ref{tab:61Co}}\\
 \hline
  E$_i$ (keV) & E$_f$ (keV) & E$_\gamma$ (keV)  & I$_\gamma$ (\%) & J$^{\pi}_i$ & J$^{\pi}_f$ & $R_{DCO}$ & Asymmetry & Multipolarity \\
 \hline
 \endhead

 \hline
 \endfoot
 
\hline
 \multicolumn{9}{| c |}{End of Table}\\
 \hline
 \endlastfoot
    1285.9(5) & 0 & 1285.9(5) & 80 (8) & 9/2$^-$ & 7/2$^-$ & 1.11 (11)$^D$ && D\\		
    1664.8(5) & 0 & 1664.8(5) & 100 (12) & 11/2$^-$ & 7/2$^-$ & 2.16 (25)$^D$ & 0.075 (12) & E2\\
	& 1285.9 & 378.4(5) & 15 (1) & 11/2$^-$ & 9/2$^-$ & 0.54 (8) && D\\	
    2339.4 (8) & 1285.9 & 1053.5(6) & 20 (2) &	11/2$^-$ & 9/2$^-$ & 0.60 (6) && D\\	
    2374.4 (8) & 1664.8 & 709.6(6) & 35 (6) & 13/2$^-$ & 11/2$^-$ & 0.66 (5) & -0.036 (11) & M1\\
    3127.0 (9) & 2374.4 &	752.6(5) &	25 (7) &	15/2$^-$ & 13/2$^-$ & 0.66 (7) & -0.041 (11) & M1\\ 
    & 1664.8 & 1462.0(6) & 14 (4) & 15/2$^-$ & 11/2$^-$ & 1.19 (12) && Q\\
    3159.4(9) &	1664.8 & 1494.6(8) & 14 (1) & 13/2$^-$ &	11/2$^-$ & 0.65 (6) && D\\	
    3472.5(7) &	1664.8 & 1807.7(5) & 42 (4) & 13/2$^+$ & 11/2$^-$ & 0.59 (6) & 0.030 (11) & E1\\
    & 2339.4 & 1131.6(6) & 8 (2) & 13/2$^+$ & 11/2$^-$ & 0.81 (8)$^D$ &0.044 (11) & E1\\
	& 2374.4 & 1097.6(9) &	3.5 (5) & 13/2$^+$ & 13/2$^-$ &&& $\Delta$J = 0 \\	
    3658.8(9) & 3472.5 & 186.3(5) & 35 (5) & 15/2$^+$ & 13/2$^+$ & 0.55 (4) && D\\	
	& 3127.0 & 531.2(5) & 15 (4) & 15/2$^+$ & 15/2$^-$ & 0.84 (8) && $\Delta$J = 0 \\	
    3910.8(7) & 1664.8 & 2246.0(5) & 4.5 (2) & 13/2$^-$ & 11/2$^-$ &&&\\		
    4094.4(10) & 3658.8 & 435.6(5) & 28 (6) & 17/2$^+$ & 15/2$^+$ & 0.68 (6) & -0.047 (11) & M1\\
    & 3127.0 & 966.6(6) & 6 (1) & 17/2$^+$ & 15/2$^-$ & 0.99 (10)$^D$ && D\\		
    4117.9(11) & 3159.4 & 956.5(7) & 12 (2) & 15/2$^-$ & 13/2$^-$	&&&\\	
    & 2374.4 & 1743.5(8) & 5 (1) & 15/2$^-$ & 13/2$^-$ & 0.40 (5) && D\\	
    & 3910.8 & 208.2(6) & 4.9 (2) & 15/2$^-$ & 13/2$^-$ & 0.66 (6) && D\\
    & 3658.8 & 456.5(7) & 4.5 (6) & 15/2$^-$ & 15/2$^+$ & & 0.108 (13) & $\Delta$J = 0\\
    4388.3(13) & 4117.9 & 270.4(7) & 12 (1) & 17/2$^-$ & 15/2$^-$ & 0.44 (4) && D\\	
    & 3658.8 & 728.3(7) & 12 (4) & 17/2$^-$ & 15/2$^+$ & 0.56 (6) && D\\	
    4485.4(8) & 3472.5 & 1012.9(3) & 12 (1) & 15/2$^-$ & 13/2$^+$ & 0.72 (8) & 0.107 (13) & E1\\
    & 3127.0 & 1358.4(6) & 8 (1) & 15/2$^-$ & 15/2$^-$ & 1.28 (11)$^D$ && $\Delta$J = 0 \\	
    4803.6(12) & 4094.4 & 709.2(6) & 27 (7) & 19/2$^+$ & 17/2$^+$ & 0.66 (5) & -0.036 (11) & M1\\
    4871.2(10) & 4485.4 & 385.8(6) & 19 (4) &	17/2$^-$ & 15/2$^-$ & 0.64 (8)	&& D\\
    & 3658.8 & 1212.4(6) & 5.0 (5) & 17/2$^-$ & 15/2$^+$ & 0.87 (12)$^D$ && D\\		
    5118.8(15) & 4388.3 & 730.5(7) & 6 (1) & 19/2$^-$ & 17/2$^-$ & 0.56 (6) && D\\	
    5347.8(12) & 4871.2 & 476.6(6) & 5.0 (5) & 19/2$^-$ & 17/2$^-$ & 0.97 (14) $^D$ && D\\		
    5723.3(13) & 3658.8 & 2064.5(9) & 7 (2) &	17/2$^-$ & 15/2$^+$ &&&\\		
    5814.3(11) & 3658.8 & 2155.5(6) & 3.4 (1) & 17/2$^-$ & 15/2$^+$ & 0.38 (5) && D\\	
    5833.1(13) & 4803.6 & 1029.5(6) & 19 (1) &	21/2$^+$ & 19/2$^+$ & 0.73 (7) && D\\	
    5955.8(17) & 5347.8 & 608.5(5) & 4.3 (4) & 21/2$^-$ & 19/2$^-$ & 0.65 (6)	&& D\\
    & 5118.8 & 837.0(8) & 3.8 (2) & 21/2$^-$ & 19/2$^-$ &&&\\		
    6066.6(17) & 5118.8 & 947.8(8) & 2.2 (4) & 21/2$^-$ & 19/2$^-$	& 0.83 (7)$^D$ && D\\	
    6170.5(12) & 4094.4 & 2076.1(6) & 10 (1) &	19/2$^-$ & 17/2$^+$ & 0.67 (9) & 0.045 (11) & E1\\
    & 5723.3 & 445.3(5) & 7 (1) & 19/2$^-$ & 17/2$^-$ &&&\\		
    6710.4(13) & 6170.5 & 539.9(6) & 5.0 (3) & 21/2$^-$ & 19/2$^-$ & 1.26 (18)$^D$ && D\\
    & 5955.8 & 754.0(9) & 4.1 (3) & 21/2$^-$ & 21/2$^-$ &&& $\Delta$J = 0 \\		
    6749.2(18) & 5955.8 & 793.4(7) & 3.4 (5) & 23/2$^-$ & 21/2$^-$ &&&\\		
    6823(13) & 6170.5 & 652.5(6) & 5.0 (5) & 21/2$^-$ & 19/2$^-$ & 0.51 (10)&& D\\		
    & 5814.3 & 1007.9(11) & 3.2 (2) & 21/2$^-$ & 17/2$^-$ & 0.96 (10) && Q\\	
    6894.9(14) & 5833.1 & 1061.8(6) & 7 (1) & 23/2$^+$ & 21/2$^+$ & 0.50 (7) && D\\	
    7506.6(15) & 6710.4 & 796.2(7) & 4.5 (2) & 23/2$^-$ & 21/2$^-$ & 0.61 (6) && D\\	
    7703.2(19) & 6749.2 & 954.0(7) & 2.8 (5) & 25/2$^-$ & 23/2$^-$ &&&\\		
    8089.5(14) & 6823 & 1266.5(6) & 3.2 (1) & 25/2$^-$ & 21/2$^-$ &0.93 (9)& 0.099 (13) & E2\\
    8216.7(16) & 6894.9 & 1321.8(8) & 4.3 (2) & 25/2$^+$ & 23/2$^+$	&&&	\\
    8408.1(16) & 7506.6 & 901.5(6) & 2.3 (2) & 25/2$^-$ & 23/2$^-$	&&&	\\
    9678.3(17) & 8216.7 & 1461.6(7) & 3.8 (3) & 27/2$^+$ & 25/2$^+$	&&&	\\

        \hline
    
\end{longtable*}

\begin{longtable*}[c]{|cccc|cccc|}
 \caption{A table showing the comparison between the energy levels from the experiment and the theoretical predictions from the shell model in $^{59}$Co, $^{59}$Ni, and $^{61}$Co. Energy levels (experiment and theory) and differences are given in keV. 2J values in parenthesis are the states with tentative assignments of parity. 2J values with ``*" indicate the second excited states (yrare states). \label{tab:Exp_Theo_Table}}\\
     \hline
     \multicolumn{8}{|c|}{Begin of Table}\\
     \hline
     &&Negative Parity States &&& Positive Parity States&&\\
     \hline
     2J & Experiment & Theory& Difference & 2J & Experiment&Theory & Difference \\
        \hline
        \endfirsthead
        
    \hline
     \multicolumn{8}{|c|}{Continuation of Table \ref{tab:Exp_Theo_Table}}\\
      \hline
     &&Negative Parity States &&& Positive Parity States&&\\
     \hline
     2J & Experiment & Theory& Difference & 2J & Experiment&Theory & Difference \\
        \hline
        \endhead
\hline
 \endfoot
 
\hline
 \multicolumn{8}{| c |}{End of Table}\\
 \hline \hline
 \endlastfoot
 \multicolumn{8}{| c |}{$^{59}$Co}\\
 \hline
7 & 0 & 0 & 0&7&-&-&-\\
9 &1191 &1342 &151 &9&-&4314&-\\
11&1460&1494&34 &11 &3844 &4497 &653\\
13&2155&2178&23 &13 &4179 &4773 &594\\
15&2915&3251&336 &15&4415&4783&368\\
(17)&3628&4756&1128 &17&4718&5000&282\\
(19)&4800&5790&990 &19&5371&5601&230\\
(21)&6879&6936&57 &21&6366&6614&248\\
(23)&7904&8337&433 &23&7460&7523&63\\
(25)&8868&9417&548 &(25)&8898&8899&1\\
(27)&9553&10506&953 &(27)&9815&9754&61\\
29&-&12338&- &29 &-&11105&-\\
31&-&13601&- &(31)&10889&12061&1172\\
11*&2184&2158&26 &&&&\\
(13)*&3225&3208&17 &&&&\\
15*&3738&4019&281 &&&&\\
(17)*&4785&4919&134 &&&&\\
(19)*&6422&6098&324 &&&&\\
21*&-&7347&- &&&&\\
(23)*&8119&8461&342 &&&&\\
25*&-&9836&- &&&&\\
27*&-&11049&- &&&&\\
29*&-&12469&- &&&&\\
31*&-&13911&- &&&&\\
         \hline
         
\multicolumn{8}{| c |}{$^{59}$Ni}\\
\hline
3 & 0 & 0&0 &9 &3057 &3267&210 \\
5 & 340 &224&116 &11&4105&4803&698 \\
7 &1339 &1189&150 &13&4458&4653&195 \\
9 &1769&1699&70 &15&5383&6050&667 \\
11 &2707&2587&120 &17&5255&5401&146 \\
13 &4144&3633&511 &19 &-&7139&- \\ 
15&4951&4469&482 &21&8133&7875&258\\ 
(17)&5948&5380&568 &23 &-&8471&- \\ 
(19) &6484&6238&246 &25&9902&9129&773\\ 
(21) &7168&7186&18 &27&-&10161&- \\ 
(23)&7954&8361&407&29&11645&11593&52 \\ 
(25)&8632&9656&1024 & 31 &-&13111&-\\
(27)&9310&11391&2081 &(33)&13229&13893 &664\\ 
(29)&10089&12951&2862 & 35&-&15946&- \\
(31)&12008&14494&2486 &(37)&15179&17047 &1868 \\                       
3*&-&966&- &39 &-&18882&- \\
5*&1189&1527&338 &(41)&17685&20283 &2598 \\
7*&1950&1749&201 &43 &-&21757&- \\
9*&2530&2652&122 &45 &-&23491&- \\
11*&3127&2883&244 &&&&\\
13*&4419&4010&409 &&&&\\
15*&5295&4735&560 &&&&\\
17*&6079&5654&425 &&&&\\
19*&6506&6419&87 &&&&\\
(21)*&7233&7999&766 &&&&\\
23*&8291&9321&1030 &&&&\\
25*&-&10587&- &&&&\\
(27)*&9753&11973&2220 &&&&\\
29*&-&13211&- &&&&\\
31*&-&14973&- &&&&\\
\hline 

\multicolumn{8}{| c |}{$^{61}$Co}\\
\hline
7 & 0 & 0 &0 &7&-&-&- \\
9 &1286 &1536&250 &9 & - &3376&- \\
11 &1665&1608&57 &11& - &3432&- \\
13 &2374&2259&115 &13&3472&3411&61 \\
15 &3127&2939&188 &15&3657 &3577&80 \\
17&4388&4833&445 &17&4093&4054&39 \\
19&5119&5898&779 &19 &4803&4932&129 \\
21&5956&7349&1393 &21&5832&5940&108 \\
23&6749&7652&903 &23&6893&6915&22 \\
25 &7703&8652&749 &25 &8135 &8140&5 \\
7*&-&1578&- &27 &9597 &9330&267 \\
9*&-&2240&- &&&&\\	
11*&2339&2447&108 &&&&\\
13*	&3159	&3138&21 &&&&\\
15*	&4118	&3747	&371 &&&&\\
17*	&4871	&5097	&226 &&&&\\
19*	&5348	&5898	&550 &&&&\\
21*	&6067	&7685	&1618 &&&&\\
23*	&7506	&8242	&736 &&&&\\
25*	&8090	&-&- &&&&\\

\end{longtable*}

\end{appendices}
\end{document}